\documentclass[fleqn,usenatbib]{mnras}

\usepackage{newtxtext,newtxmath}

\usepackage[T1]{fontenc}

\DeclareRobustCommand{\VAN}[3]{#2}
\let\VANthebibliography\thebibliography
\def\thebibliography{\DeclareRobustCommand{\VAN}[3]{##3}\VANthebibliography}

\usepackage{graphicx}	
\usepackage{amsmath}	

\usepackage{amssymb}	
\usepackage{cite}

\title[Neutrino Cosmology from CSST Photometric Surveys]{Forecast of Neutrino Cosmology from the CSST Photometric Galaxy Clustering and Cosmic Shear Surveys}

\author[H. Lin et al.]{
Hengjie Lin$^{1,2}$,
Yan Gong$^{1,3}$\thanks{E-mail: gongyan@bao.ac.cn},
Xuelei Chen$^{4, 2, 5}$,
Kwan Chuen Chan$^{6,7}$,
Zuhui Fan$^8$,
Hu Zhan$^{1,9}$
\\
$^{1}$Key Laboratory of Space Astronomy and Technology, National Astronomical Observatories, Chinese Academy of Sciences, \\20A Datun Road, Beijing 100012,
China\\
$^{2}$School of Astronomy and Space Sciences, University of Chinese Academy of Sciences, Beijing 100049, China\\
$^{3}$Science Center for China Space Station Telescope, National Astronomical Observatories, Chinese Academy of Sciences, \\20A Datun Road, Beijing 100101, China\\
$^{4}$Key Laboratory for Computational Astrophysics, National Astronomical Observatories, Chinese Academy of Sciences, \\20A Datun Road, Beijing 100101, China \\
$^{5}$Centre for High Energy Physics, Peking University, Beijing 100871, China \\
$^{6}$School of Physics and Astronomy, Sun Yat-sen University, 2 Daxue Road, Tangjia, Zhuhai, 519082, China  \\
$^{7}$CSST Science Center for the Guangdong-Hongkong-Macau Greater Bay Area, SYSU,  Zhuhai, 519082, China \\
$^{8}$South-Western Institute for Astronomy Research, Yunnan University, Kunming 650500, China \\
$^{9}$Kavli Institute for Astronomy and Astrophysics, Peking University, Beijing 100871, China \\
}

\date{Accepted XXX. Received YYY; in original form ZZZ}

\pubyear{2015}

\begin{document}
\label{firstpage}
\pagerange{\pageref{firstpage}--\pageref{lastpage}}
\maketitle

\begin{abstract}
China Space Station Telescope (CSST) is a forthcoming powerful Stage IV space-based optical survey equipment. It is expected to explore a number of important cosmological problems in extremely high precision. In this work, we focus on investigating the constraints on neutrino mass and other cosmological parameters under the model of cold dark matter with a constant equation of state of dark energy ($w$CDM), using the mock data from the CSST photometric galaxy clustering and cosmic shear surveys (i.e. 3$\times$2pt). The systematics from galaxy bias, photometric redshift uncertainties, intrinsic alignment, shear calibration, baryonic feedback, non-linear, and instrumental effects are also included in the analysis. We generate the mock data based on the COSMOS catalog considering the instrumental and observational effects of the CSST, and make use of the Markov Chain Monte Carlo (MCMC) method to perform the constraints. Comparing to the results from current similar measurements, we find that CSST 3$\times$2pt surveys can improve the constraints on the cosmological parameters by one order of magnitude at least. We can obtain an upper limit for the sum of neutrino mass $\Sigma m_{\nu} \lesssim 0.36$ (0.56) eV at 68\% (95\%) confidence level, and $\Sigma m_{\nu} \lesssim 0.23$ (0.29) eV at 68\% (95\%) confidence level if ignore the baryonic effect, which is comparable to the {\it Planck} results and much better than the current photometric surveys. This indicates that the CSST photometric surveys can provide stringent constraints on the neutrino mass and other cosmological parameters, and the results also can be further improved by including data from other kinds of CSST cosmological surveys.
\end{abstract}




\section{Introduction}

Studies of the physical nature of dark energy and dark matter are the key topics in modern cosmology. Current cosmological  observations, for example, Type Ia supernovae \citep[e.g.][]{scolnic2018IaSNe, Riess-2021IaSNe}, cosmic microwave background (CMB) \citep[e.g.][]{hinshaw2013wmap, Planck2018-I} and baryon acoustic oscillations (BAO) \citep[e.g.][]{alam2021eBOSS}, suggest that our Universe is flat and with about $30\%$ matter (visible matter and dark matter) and $70\%$ dark energy. Among these components, the neutrino is a major topic in both modern cosmology and fundamental particle physics, which can act as hot dark matter in the Universe. It can significantly affect the formation and evolution of cosmic large-scale structure (LSS), especially at relatively small scales, and imprint remarkable features in galaxy formation and clustering.

Currently, the ground based laboratory neutrino oscillation experiments can only constrain the mass-squared differences $\Delta m_{21}^2$ and $\lvert \Delta m_{31}^2 \rvert$, and then further evaluate possible range of total neutrino mass $\Sigma m_{\nu}$. According to the sign of $\Delta m_{31}^2$, we have two types of mass orderings, i.e. the normal hierarchy ($\Delta m_{31}^2 > 0$) and the inverted hierarchy ($\Delta m_{31}^2 < 0$). The lower limit of total neutrino mass can be found as $\Sigma m_{\nu} \gtrsim 0.06 \text{ eV}$ for the normal hierarchy or $\Sigma m_{\nu} \gtrsim 0.1 \text{ eV}$ for the inverted hierarchy \citep{Amsler2008}. On the other hand, cosmological observations can provide an effective way to constrain the total neutrino mass. For example, the precise measurements of cosmic microwave background (CMB) by the $\it Planck$ satellite give an upper limit $\Sigma m_{\nu} \lesssim 0.12 \text{ eV}$ by combining {\it Planck} 2018 data and BAO measurement \citep{Planck2018-VI}. If more observational data are included, it might be able to improve the constraint power even further. For example, in \citet{Giusarma-mnu} and \citet{Vagnozzi-mnu}, they combined BAO measurements and galaxy power spectra from different galaxy surveys, local measurements of the Hubble parameter, the measurement of the optical depth to reionization by the Planck satellite, cluster counts from the Planck satellite thermal Sunyaev-Zeldovich (SZ) effect, and the CMB data by {\it Planck} 2015. They obtained similar constraint power on neutrino mass with the {\it Planck} 2018 result about one year earlier, showing the promising potential of the joint constraint power. Those constraints are close to the minimum total mass derived in the inverted hierarchy, so it will be quite helpful to include cosmological data for distinguishing the neutrino mass hierarchies.

Besides CMB observations related to the early universe, measurements of cosmic large-scale structure at lower redshifts also can provide powerful tools to extract properties of neutrinos. As we discuss in the next section, neutrinos can significantly suppress matter density fluctuations at small scales, and can affect the formation and clustering of galaxies. Hence, we can directly measure galaxy clustering to derive the properties of neutrinos. 
To obtain accurate information of galaxy clustering, spectroscopic surveys are usually performed, which can get precise galaxy redshifts, such as 2dF \citep[e.g.][]{Colless-2df, Cole-2df}, 6dF \citep{Beutler-6df}, WiggleZ \citep[e.g. ][]{Blake-WiggleZ, Parkinson-WiggleZ}, and Baryon Oscillation Spectroscopic Survey (BOSS) \citep{Dawson-BOSS, Dawson-eBOSS} surveys. On the other hand, photometric surveys also can be used to measure the LSS. Although the redshift information is not as accurate as in spectroscopic surveys, photometric surveys can provide much larger and deeper galaxy samples, which can precisely measure the two dimensional (2D) galaxy clustering information integrated over large redshift range. In addition, it also can easily cross-correlate with cosmic shear measurements to further improve the cosmological constraints and reduce relevant systematical uncertainties.

Cosmic shear surveys are another powerful photometric probe of measuring the LSS \citep{kaiser1992,Waerbeke2000}. By detecting galaxy shear signals or weak gravitational lensing effect caused by foreground gravitational potentials of massive objects, unbiased measurements of matter density fluctuations in the Universe can be obtained. Nowadays, there are a number of optical surveys that are dedicated to obtaining measurements of the weak lensing signal, such as Dark Energy Survey (DES) \citep[e.g.][]{Abbott2018DES, DES2021}, Kilo-Degree Survey (KiDS) \citep[e.g.][]{van2018kids, Heymans-KiDS}, Subaru Hyper Suprime-Cam (HSC) \citep[e.g.][]{hikage2019HSC, Hamana-HSC}. These surveys have provided effective constraints on several important cosmological parameters related to dark energy, dark matter, and neutrinos. 

The upcoming Stage IV surveys, such as Large Synoptic Survey Telescope (LSST) \citep{ivezic2009lsst} or Vera C. Rubin Observatory, {\it Euclid} \citep{laureijs2011euclid}, Nancy Grace Roman Space Telescope (RST) or Wide-Field Infrared Survey Telescope (WFIRST) \citep{green2012wfirst}, and China Space Station Telescope (CSST) \citep{zhan2011csst, zhan2018csst, zhan2021csst, Gong-CSST-2019} are aiming to achieve much tighter constraints on the key cosmological parameters by using photometric and spectroscopic probes we mention above. It will take a great step of understanding the nature of the dark sector, and might have a great chance to determine the hierarchy of neutrino mass. 

In this work, we focus on the CSST photometric surveys to study the constraints on neutrino mass and other cosmological parameters. The CSST is a 2-meter space telescope in the same orbit of the China Manned Space Station. The CSST survey will cover 17,500 deg$^2$ sky area with the field of view 1.1 deg$^2$. It will carry multiple scientific equipments, that allows it to collect photometric image and spectroscopic data in the mean time. It has seven photometric and three spectroscopic bands from near-UV to near-IR covering 255-1000 nm, i.e. $NUV$, $u$, $g$, $r$, $i$, $z$, and $y$ bands for photometric surveys, and $GU$, $GV$, and $GI$ bands for spectroscopic surveys. The magnitude limit of the CSST photometric survey can reach $i\simeq26$ AB mag for 5$\sigma$ point source detection. The CSST survey will be a powerful survey for probing the Universe and studying lots of important cosmological and astronomical topics. Here we mainly explore the prediction of constraints on neutrino cosmology in the CSST photometric galaxy clustering and cosmic shear surveys. We generate mock data of CSST galaxy angular power spectrum, shear power spectrum, and galaxy-galaxy lensing power spectrum (i.e. 3$\times$2pt) in tomographic redshift bins from z=0 to 4, and consider possible instrumental and astrophysical systematics, such as the uncertainties from photometric redshift (photo-$z$), galaxy bias, point-spread function (PSF), intrinsic alignment, baryonic feedback, non-linear effect, and so on. Then we perform a joint fit simultaneously including both cosmological and systematical parameters with all 3$\times$2pt data to study the constraint results of neutrino cosmology.

The paper is organized as follows: in Section~\ref{sec:Neu_cosmos}, we introduce neutrino cosmology and the effects of massive neutrinos on the matter density fluctuations; Section~\ref{sec:mock_data} presents the details of theoretical models we adopt to predicate the galaxy clustering and cosmic shear power spectra, and generation of the 3$\times$2pt mock data including relevant systematics; The model fitting method and the discussion of constraint result of cosmological and systematical parameters are given in Section~\ref{sec:results}. Finally, we summarize the conclusion and give relevant discussions in Section~\ref{sec:Conclusions}.

\begin{figure*}
    \includegraphics[scale = 0.33]{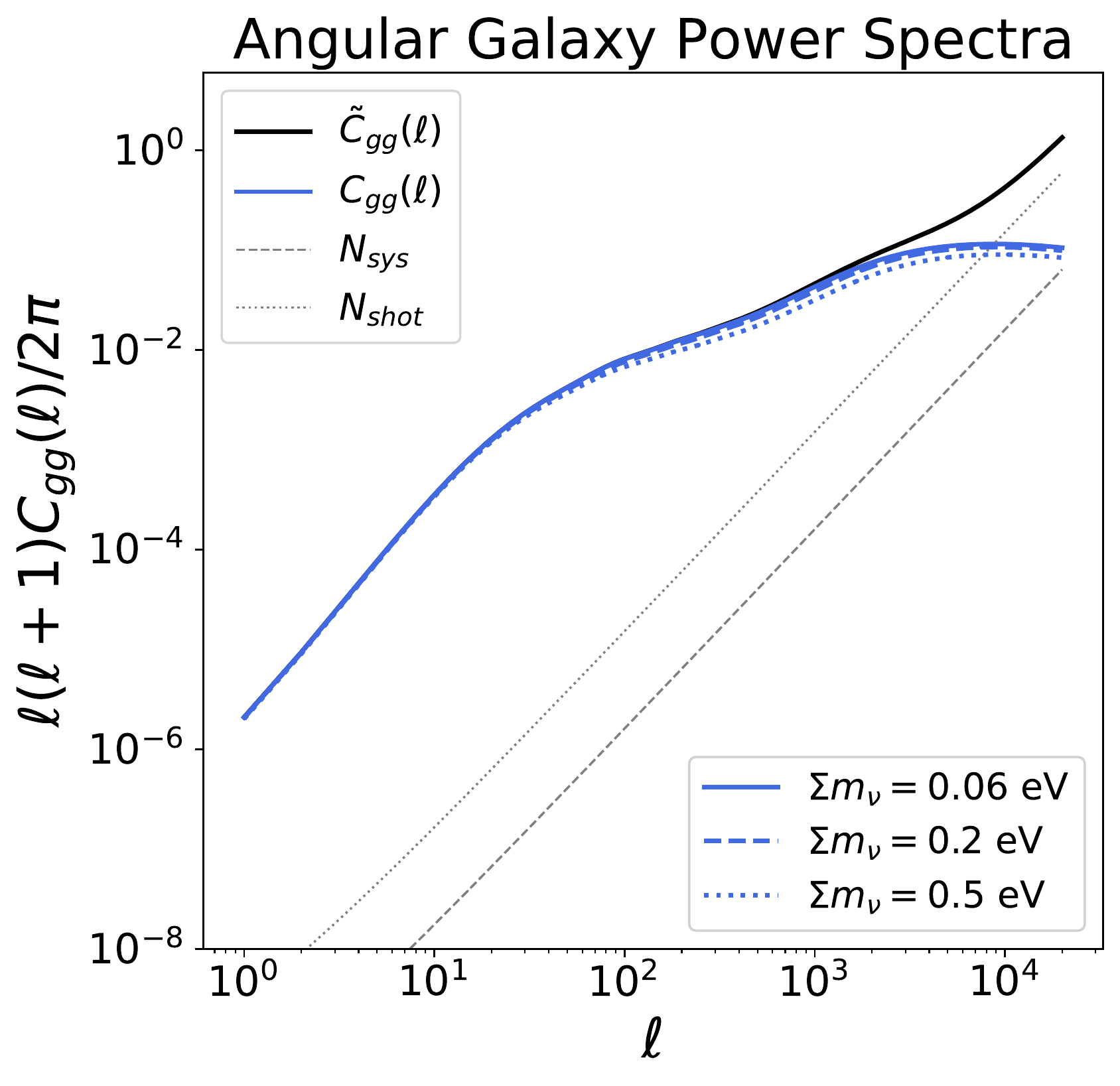}
    \includegraphics[scale = 0.33]{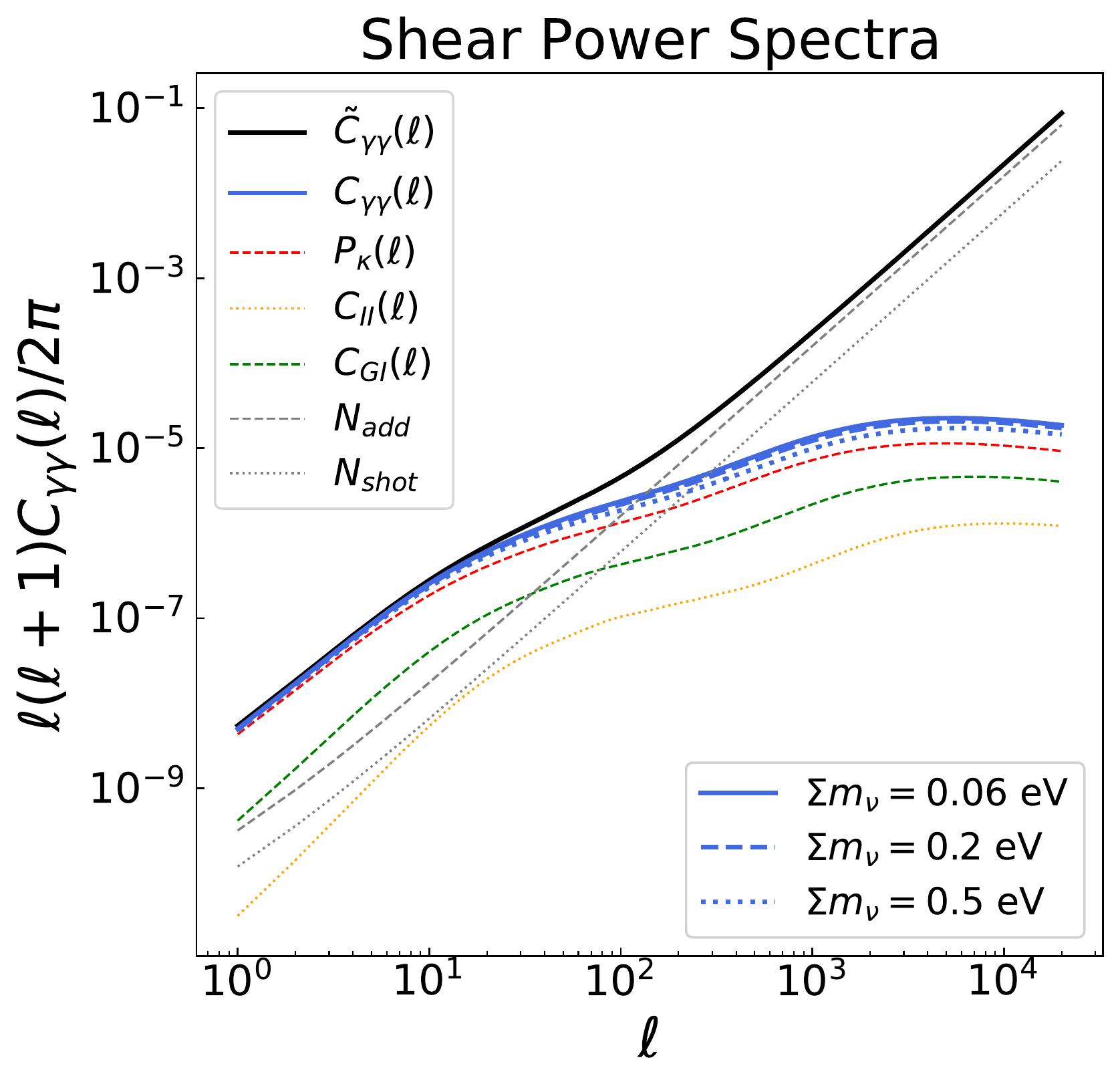}
    \includegraphics[scale = 0.33]{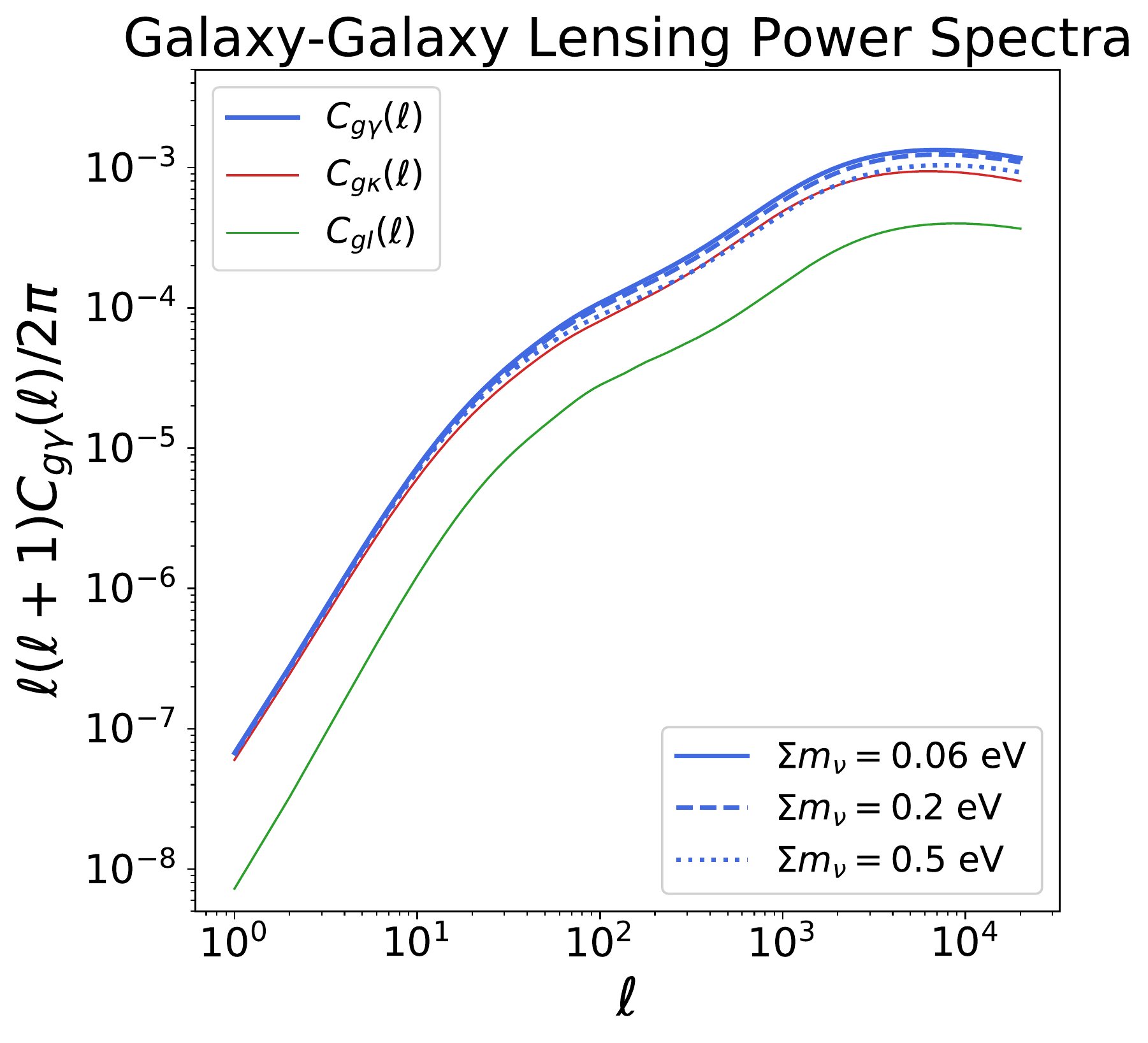}
    \caption{{\it Left panel}: the components of the angular galaxy power spectrum. The black solid, blue solid, gray dotted and gray dashed curves denote the total, signal, shot-noise and systematical noise power spectra, respectively. {\it Middle panel}: the components of the cosmic shear power spectrum. The black solid, blue solid, red dashed, orange dotted, and green dashed curves denote the total, total signal, convergence, intrinsic-intrinsic, gravitational-intrinsic power spectra. The shot-noise and additive noise are presented by gray dotted and gray dashed curves. {\it Right panel}: the components of galaxy-galaxy lensing power spectrum. The blue, red, and green solid curves are total, galaxy-galaxy lensing signal, and galaxy-intrinsic alignment power spectra. We also show the suppression of the neutrino free-streaming effect on the power spectra in all panels. The blue solid curves stand for the fiducial cosmology with $\Sigma m_{\nu} = 0.06$ eV, and the blue dashed and dotted curves denote the $\Sigma m_{\nu} = 0.2$ eV and $0.5$ eV cases.}
    \label{fig:shear-with-components}
\end{figure*}

\section{Neutrino cosmology}\label{sec:Neu_cosmos}

\subsection{Massive Neutrino}\label{subsec:massive_neutrino}

Massive neutrinos can impact the formation and evolution of the cosmic large-scale structure through the free streaming process \citep[e.g.][]{Bond-free-streaming}. Massive neutrinos will suppress the matter clustering on scales smaller than a comoving free-streaming scale, or, in $k$-space, the wavenumbers $k > k_{\text{FS}}$, and thus imprint relevant information of neutrino properties in the matter power spectrum. The suppression scale depends on the thermal velocity of non-relativistic neutrinos $v_{\rm t}$, and it can be estimated by
\begin{equation}
k_{\text{FS}}(z)=\sqrt{\frac{3}{2}} \frac{H(z)}{v_{\rm t}(z)(1+z)}\ h\ \text{Mpc}^{-1}.
\label{eq:k_FS}
\end{equation}
Here $H(z)$ is the Hubble parameter. In the flat $w$CDM universe including non-relativistic neutrinos and ignoring radiation contribution, it can be expressed as
\begin{equation}
H(z) = H_0\sqrt{\Omega_{\rm m}(1+z)^3+\Omega_{\rm DE}(1+z)^{3(1+w)}},
\end{equation}
where $H_0=100\, h\ \rm km s^{-1} Mpc^{-1}$ is the Hubble constant, $\Omega_{\rm DE}=1-\Omega_{\rm m}$ in a flat universe, $w$ is the equation of state of dark energy, and $\Omega_{\rm m}=\Omega_{\rm c}+\Omega_{\rm b}+\Omega_{\nu}$ is the present total matter energy density parameter, which is the sum of cold dark matter, baryons, and non-relativistic neutrinos. We take the present neutrino energy density parameter as
\begin{equation}
\Omega_{\nu}\simeq \frac{\sum m_{\nu}}{94.1h^2\,\rm eV},
\end{equation}
where $\sum m_{\nu}$ is the total neutrino mass. The thermal velocity of non-relativistic neutrinos can be evaluated by
\begin{equation}
v_{\rm t}(z) \simeq \frac{3.15T_{\nu}}{m_{\nu}} \simeq 158(1+z)\left(\frac{1\rm eV}{m_{\nu}}\right)\ \rm km\,s^{-1}.
\end{equation}
Here $T_{\nu}=(4/11)^{1/3}T_{\gamma}$ is the neutrino temperature, and $T_{\gamma}=T_{\gamma}^0(1+z)$ is the photon temperature, where $T_{\gamma}^0=2.726$ K. In Eqation~\ref{eq:k_FS}, we can find that, when neutrinos become non-relativistic during matter domination era, which is usually the case for neutrinos, the comoving free-streaming scale is always decreasing in time. This indicates that it exists a maximum free-streaming scale at the time of the non-relativistic transition $z_{\rm nr}$, i.e. when $m_{\nu}=3.15T_{\nu}$. Since $T_{\nu}=(4/11)^{1/3}T_{\gamma}=1.68\times10^{-4}(1+z)\ \rm eV$, we have $1+z_{\rm nr}\simeq 1900(m_{\nu}/{\rm eV})$. Hence we can find that
\begin{equation}
k_{\rm nr}=k_{\rm FS}(z_{\rm nr}) \simeq 0.018\, \Omega_{\rm m}^{1/2} \left(\frac{m_{\nu}}{1\,\rm eV}\right)^{1/2}\ h\, {\rm Mpc^{-1}}.
\end{equation}
The free-streaming scales at lower redshifts are always smaller than this scale, which means the thermal velocity of neutrino is always less than the escape velocity of gravitational potential at $k\ll k_{\rm nr}$, i.e. neutrinos cannot suppress matter density fluctuations and act as cold dark matter at these scales. 
At $k\gg k_{\rm nr}$, for neutrinos with low mass, a good approximation of the suppression of the current matter power spectrum is found to be \citep{Hu&Eisenstein-neutrino}
\begin{equation}
\frac{\Delta P_{\rm m}}{P_{\rm m}} \simeq -8f_{\nu},
\end{equation}
where $f_{\nu}=\Omega_{\nu}/\Omega_{\rm m}$. Note that this relation is only available when $f_{\nu}\lesssim0.07$ or $\sum m_{\nu}\lesssim 1\ \rm eV$ \citep[][]{Brandbyge-2008, Bird-2012}, and more accurate result can be obtained by solving the Boltzmann equation of cosmological perturbations numerically. In this work, we adopt the Boltzmann code $\tt CAMB$  \citep{CAMB-Lewis} and $\tt HMcode$ \citep{Mead-2021} to calculate the matter power spectrum including massive neutrinos and non-linear effects.

Since massive neutrino can damp the matter density fluctuations, it will affect matter gravitational potential well and hence galaxy formation and clustering. Therefore, we can explore galaxy clustering and cosmic shear to derive the properties of massive neutrinos. In Figure~\ref{fig:shear-with-components}, we show the photometric angular galaxy power spectrum, shear power spectrum, and their cross-correlation, i.e. galaxy-galaxy lensing power spectrum, with different total neutrino mass $\sum m_{\nu}$ considered (in blue solid, dashed, and dotted curves for $\sum m_{\nu}=0.06$, 0.2, and 0.5 eV, respectively). We can see that neutrinos obviously can suppress these power spectra at $\ell\gtrsim 100$, and accurate measurements of them can provide effective constraints on neutrino properties.

\subsection{Baryonic Effect}\label{subsec:baryonic_effect}

As we have discussed above, in order to extract the information that massive neutrino imprint on the LSS, we need to explore the small scales which are dominated by non-linear physics. A major systematic known as baryonic effect (BE) can have a large impact in the non-linear region.
The baryonic effect has two main processes. First, the contraction of radiative cooling gas will alter the dark matter distribution via gravitational force, causing the total matter distribution changes \citep{Duffy-2010}; Second, the violent energy released by supernova explosions or active-galactic nuclei (AGN) can heat gas and push it to outskirt of dark matter halo \citep{Schaye-2010, Chisari-2018, vanDaalen-2020}. The detail of those mechanisms still are not well-understood, but it has been well-demonstrated by high-resolution hydrodynamic simulations. So a common way to study the impact of baryonic effect is to fit the parameterized baryonic model to match the data from hydrodynamic simulations, such as COSMO-OWLS \citep{COSMO-OWLS} and BAHAMAS \citep{BAHAMAS}.

A general treatment is to introduce two parameters, one to describe the increased concentration in the halo core caused by the gas cooling process, and another one to describe the puffing up of halo profile arising from the AGN (or supernova) feedback. In \citet{Mead-2015}, they use parameter $A$ to modify concentration relation of haloes and $\eta$ to change halo profile in a mass dependent way. In \citet{Mead-2020} they consider a more complex six-parameters model, which contains a halo concentration parameter $B$; an effective halo stellar mass fraction $f_{*}$, in order to model the power spectra of stellar matter; a halo mass threshold $M_{\text{b}}$ for taking account of gas expulsion effect. They also consider the redshift evolution of those three parameters above, modeled by $B_{z}$, $f_{*,z}$ and $M_{\text{b},z}$, so it would be 6 free parameters in total.
But recent study \citep{Mead-2021} pointed out, there is some degeneracy between those two (or six) parameters in previous models, and all fitted parameters can be related to one single parameter, i.e. $\text{log}_{10}(T_{\text{AGN}}/\text{K})$, named "AGN temperature". We should be aware that $T_{\text{AGN}}$ is not a real physical observable that we can measure, and it is just a parameter to describe the strength of feedback. 

In order to get a better constraint on neutrino masses, we need to investigate whether baryonic effects are robustly modeled. In \citet{Mead-2021}, they fit their 6-parameters and single-parameter feedback model to match the data from BAHAMAS and COSMO-OWLS hydrodynamical simulations with different cosmologies. And they obtained that, for the BAHAMAS simulations, all 6 parameters of the previous version can be linearly fitted as functions of $\text{log}_{10}(T_{\text{AGN}}/\text{K})$, i.e. $P_{\text{b}}^{i} = \alpha^{i} + \beta^{i}\theta$, where $P^i_{\text{b}}$ is the $i$th parameter of the six baryonic parameters described above, $\theta \equiv \text{log}_{10}(T_{\text{AGN}}/10^{7.8}\text{K})$, and $\alpha^i$ and $\beta^i$ are the fitting parameters corresponding to the $i$th baryonic parameter. That means the single-parameter model is able to completely recover the fitting result of the 6-parameter model. They also fit the single-parameter model in the simulations based on {\it WMAP} 9 and {\it Planck} 2015 cosmologies with the existence of massive neutrinos, where the mass range of neutrino is $0.06 \leqslant \Sigma m_{\nu} \leqslant 0.48$. Although this mass range is smaller than the prior we set for neutrino masses in our forecast, we assume that a slight extension of the mass range would not significantly decrease the accuracy of the model. We also notice that, the 1$\sigma$ upper limit of neutrino mass given by our constraint results (details will be discussed in \ref{subsec:Neutrino-Mass}), are within the mass range of those simulations we mentioned above.
Therefore, we will use the single-parameter baryonic feedback model, which can avoid the degeneracies in the 6-parameter model, and can speed up the MCMC fitting process as well. We set the fiducial value as $\text{log}_{10}(T_{\text{AGN}}/\text{K}) = 7.8$, since this value can reproduce a simulation result which has good agreement with the observed galaxy stellar mass function and the hot gas mass fractions\citep{BAHAMAS}. We also adopt the prior range of $7.4 < \text{log}_{10}(T_{\text{AGN}}/\text{K}) < 8.3$ which is recommended by \citet{Mead-2021}.

\begin{figure*}
	\includegraphics[scale = 0.6]{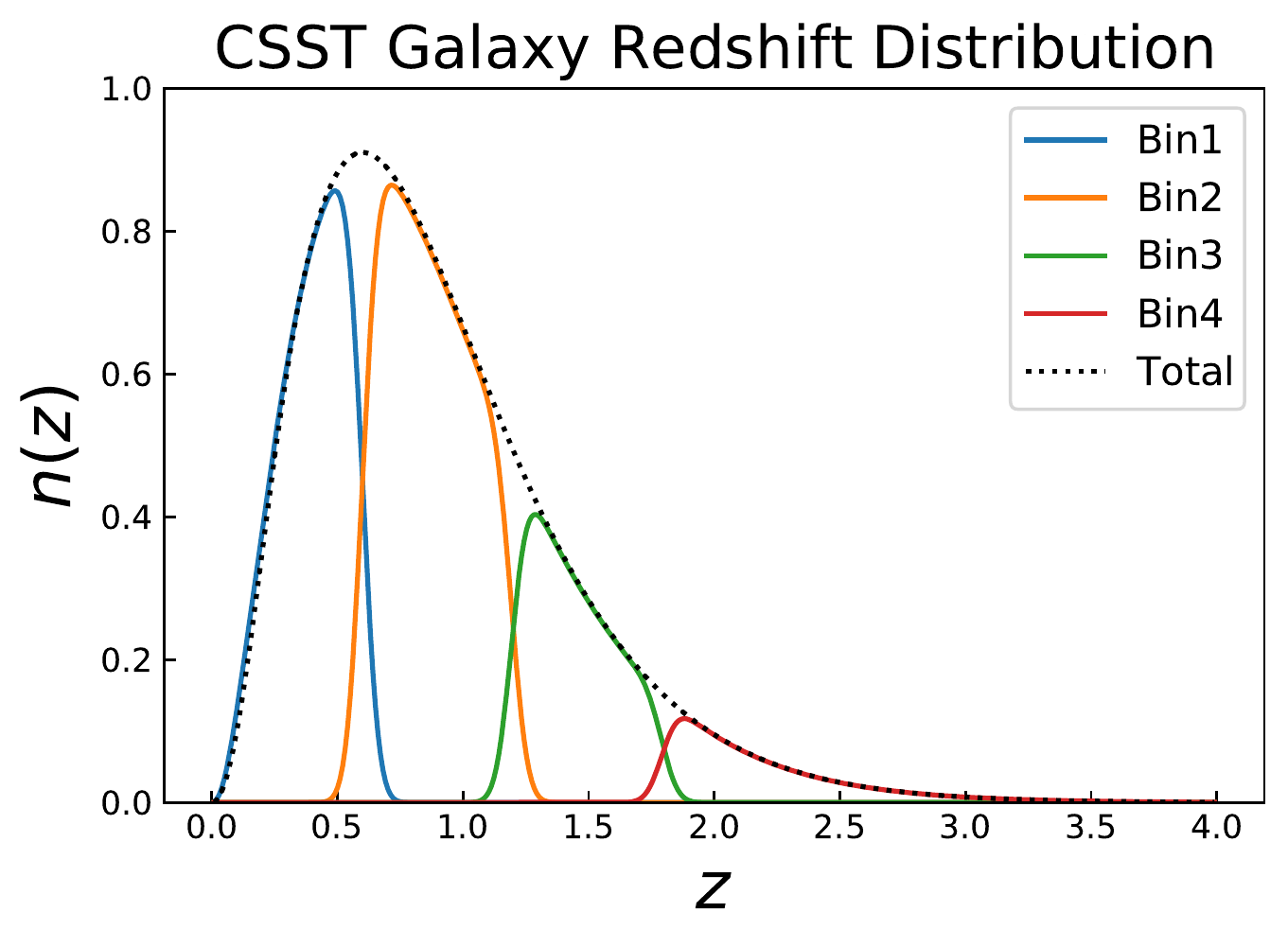}
	 \includegraphics[scale = 0.613]{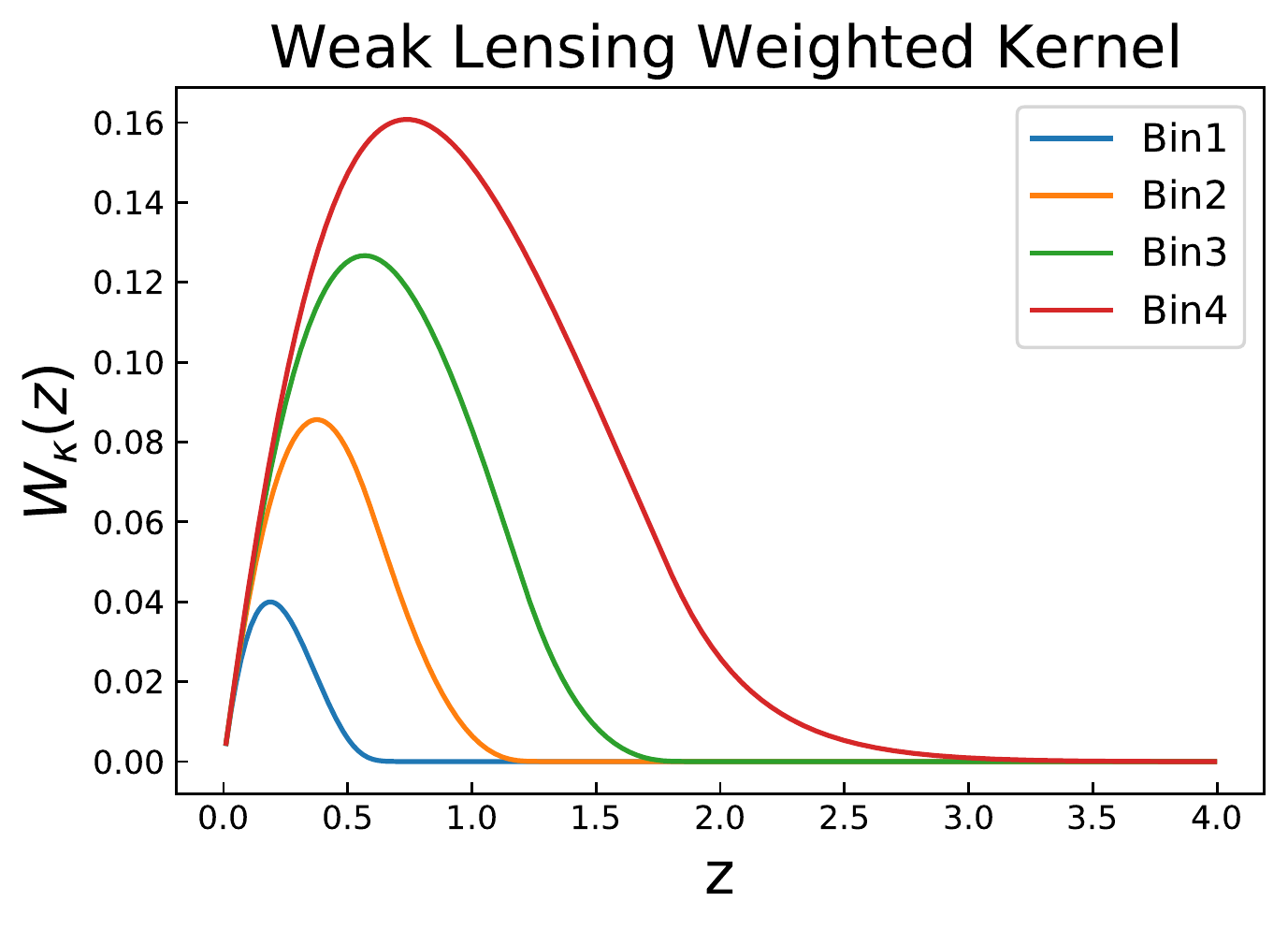}
    \caption{{\it Left panel}: galaxy redshift distribution of the CSST photometric surveys. The black dotted curve shows the total redshift distribution $n(z)$, and the solid blue, orange, green and red curves show the $n_i(z)$ for the four tomographic bins, with $\Delta_z = 0$ and $\sigma_z = 0.05$. {\it Right panel}: the weighting kernels of the four tomographic bins in the CSST cosmic shear survey.}
    \label{fig:z-dist}
\end{figure*}

\begin{figure}
    \centering
    \includegraphics[scale = 0.57]{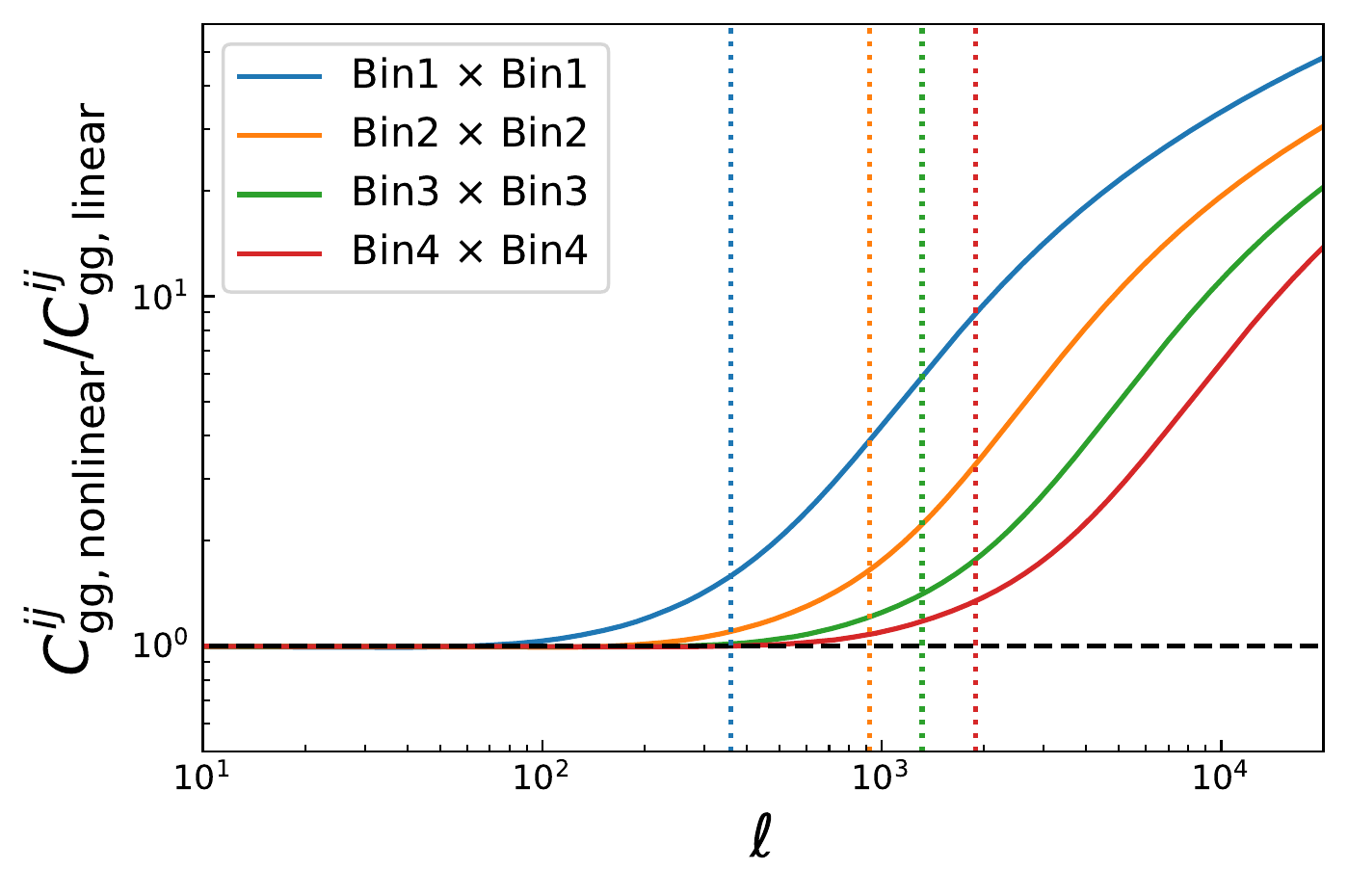}
    \caption{The ratio of non-linear to linear auto galaxy angular power spectra of the four tomographic bins shown as the solid blue, orange, green and red curves. The corresponding $\ell_{\text{max}}$ of the given upper limit $k_{\text{max}} < 0.3 h$ Mpc$^{-1}$ are denoted in dotted vertical lines.}
    \label{fig:scale-cut}
\end{figure}

\begin{figure*}
	\includegraphics[scale = 0.52]{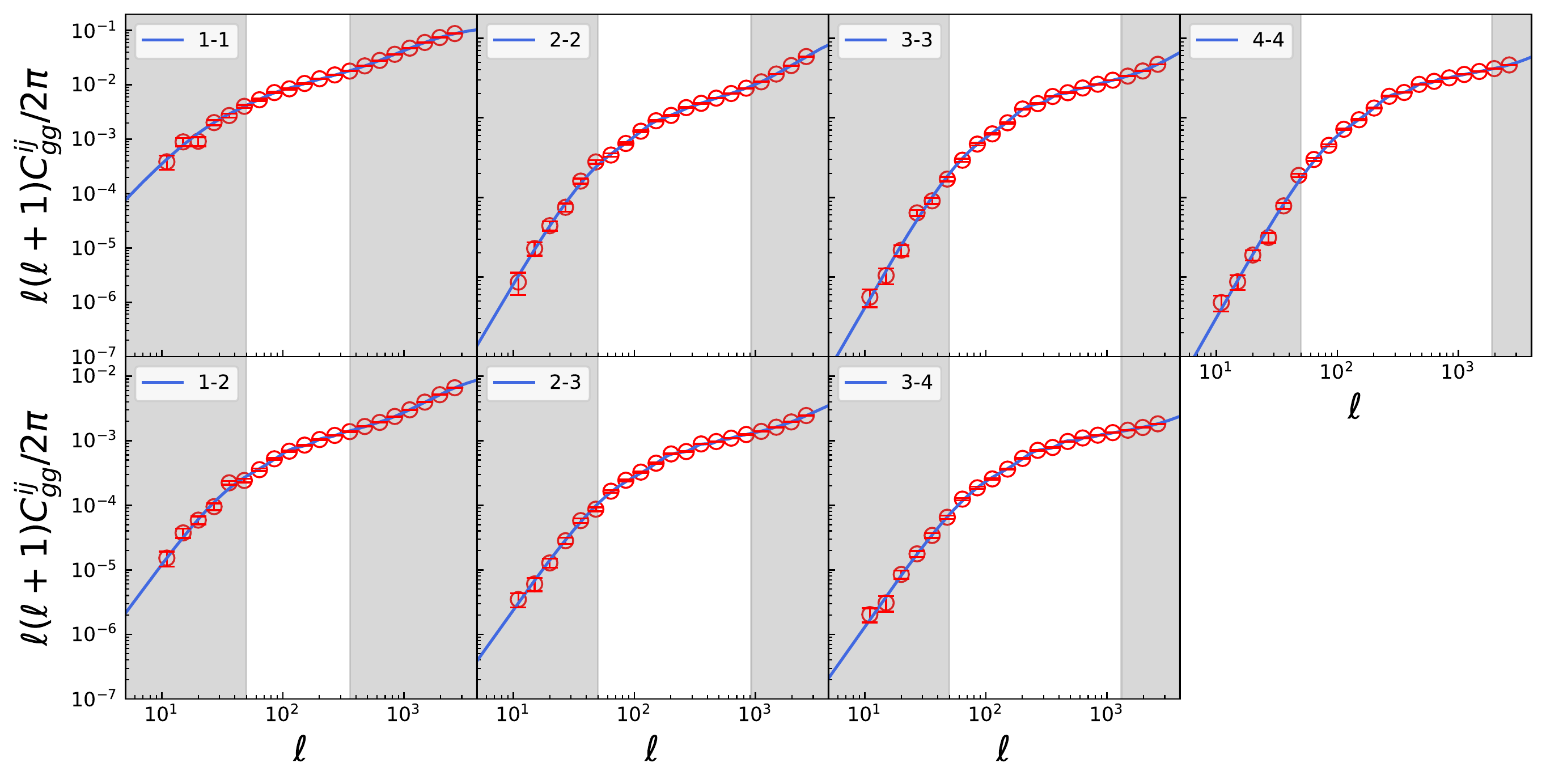}
    \caption{The mock CSST angular galaxy power spectra of the four tomographic bins. The blue solid curves show the results of the fiducial theoretical model, and red data points are the mock data. Since the cross power spectra between nonadjacent bins are quite small, we only consider three cross power spectra between adjacent bins with significant amplitudes. The gray regions show the scales for excluding the Limber approximation and non-linear effects.}
    \label{fig:gg_power_spectrum}
\end{figure*}

\section{Mock Data}\label{sec:mock_data}

\begin{figure*}
	\includegraphics[scale = 0.52]{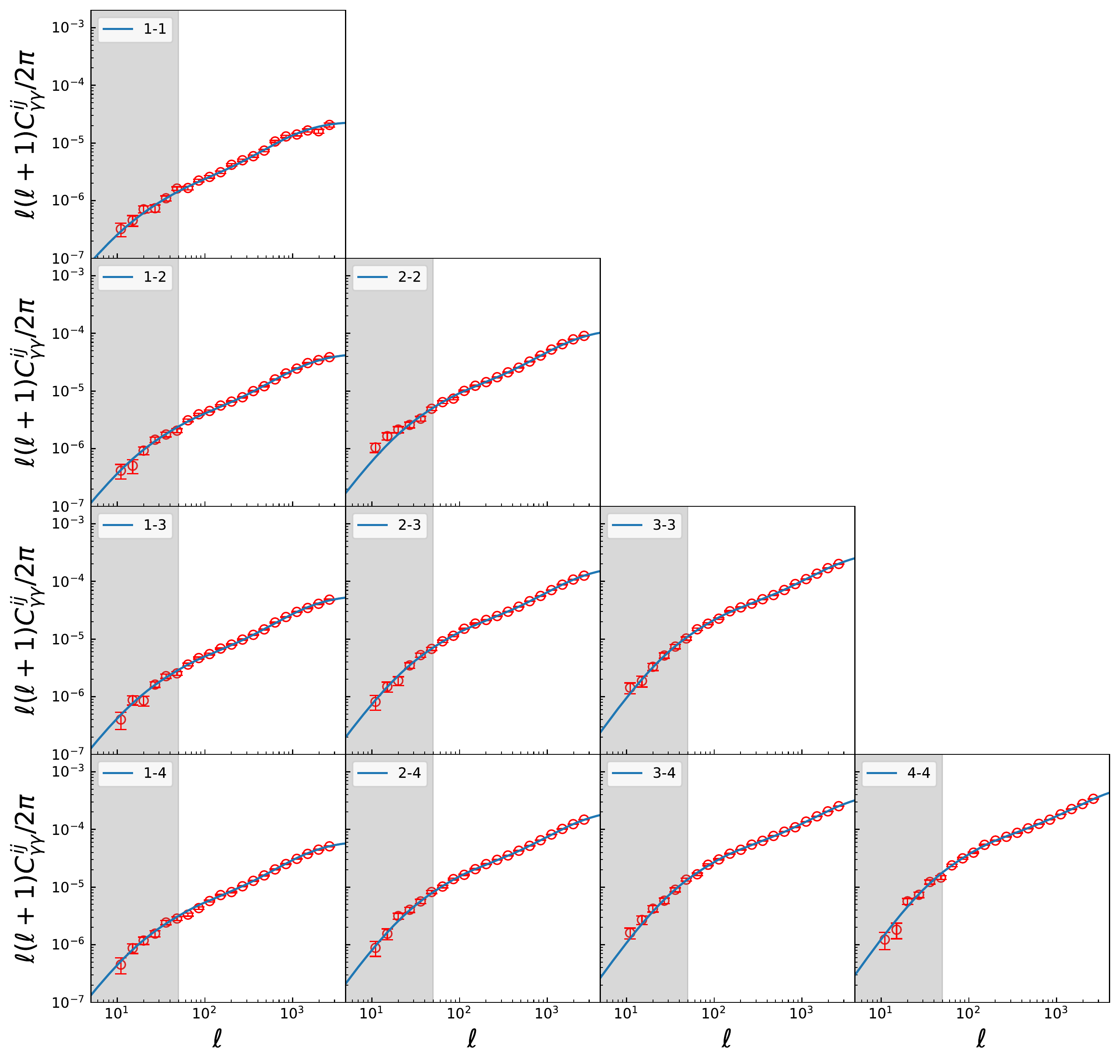}
    \caption{The mock auto and cross signal power spectra for the four tomographic bins in the CSST cosmic shear survey. The blue curves and red data points denote the theoretical models and mock data. The gray regions exclude the scales at $\ell<50$ where the Limber approximation may not be available.}
    \label{fig:shear_power_spectrum}
\end{figure*}

\subsection{Galaxy Photometric Redshift Distribution}\label{subsec:CS}

In order to generate the mock 3$\times$2pt data of the CSST photometric surveys, first we need to find the galaxy redshift distribution. Here we adopt the distribution suggested in \citet{Cao2018}. This galaxy redshift distribution is derived based on the COSMOS catalog  \citep{COSMOS1, COSMOS2}. This catalog has similar magnitude limit as the CSST photometric survey with $i\le 25.2$ for galaxy detection, and contains about 220,000 sources in a 1.7 deg$^2$ field. As shown in \citet{Cao2018}, the galaxy redshift distribution of the CSST photometric survey has a peak around $z=0.6$, and the range can extend to $z\sim4$. For analytical use, here we simply use a smooth function to represent this redshift distribution, which takes the form as
\begin{equation}
    n(z) \propto z^2e^{-z/z^{*}},
    \label{eq:z-dist-2}
\end{equation}
where $z^* = z_{\rm peak}/2 = 0.3$ in our case. In the left panel of Figure~\ref{fig:z-dist}, we show the normalized $n(z)$ with $\int n(z){\rm d}z=1$ in black dotted curve. By comparing to the distribution given in \citet{Cao2018}, we can find that they can be well matched. Besides, to extract more information from the photometric data, as shown by the solid curves, we divide the redshift distribution in several tomographic bins, and will measure the auto and cross galaxy clustering or shear power spectra for these bins. We assume the photo-$z$ bias $\Delta_z=0$, and photo-$z$ scatter $\sigma_z=0.05$, and the details of division method can be found in \citet{Gong-CSST-2019}. Here we divide $n(z)$ into four tomographic bins, which is the same as that in \citet{Gong-CSST-2019}, and this will make it easy to compare our constraint results with theirs. Note that more tomographic bins basically can further improve the constraint result, and we can use more bins in the real data analysis. For example, six photo-$z$ bins can improve the constraints by a factor of $\sim1.5$ \citep{Gong-CSST-2019}. In the current stage, four tomographic bins are sufficient for this work.

\subsection{Galaxy Angular Power Spectrum}\label{subsec:GC}

The CSST photometric galaxy survey is expected to observe billions of galaxies in the range of $z=0\sim4$, and the surface galaxy density can reach $\sim28$ arcmin$^{-2}$ \citep{Gong-CSST-2019}. It could precisely illustrate the angular distribution of galaxy clustering and map the LSS in wide fields over a large redshift range. The observed auto or cross galaxy angular power spectrum between the $i$th and $j$th bins can be written as
\begin{equation}
    \Tilde{C}_{\text{gg}}^{ij} = C_{\text{gg}}^{ij} + \frac{\delta_{ij}}{\Bar{n}_i} + N_{\text{sys}}^{\text{g}},
    \label{eq:C_gg-2}
\end{equation}
where $\Tilde{C}_{\text{gg}}^{ij}$ is the angular signal power spectrum. Assuming Limber approximation \citep{Limber}, it can be calculated by \citep{Hu&Jain2004}
\begin{equation}
    C_{\text{gg}}^{ij}(\ell) = \frac{1}{c} \int dz H(z)D_{\text{A}}^{-2}(z)W_{\text{g}}^{i}(z)W_{\text{g}}^{j}(z)P_{\text{m}}\left(\frac{\ell + 1/2}{D_{\text{A}}(z)}, z\right).
    \label{eq:C_gg-1}
\end{equation}
Here $c$ is the speed of light, $D_{\text{A}}$ is the comoving angular diameter distance, $P_{\rm m}$ is the matter power spectrum, and $W_{\text{g}}^{i}$ is the weighted integral kernel in the $i$th bin, which takes the form as
\begin{equation}
    W_{\text{g}}^i(z) = b(z) n_i(z),
    \label{eq:gg_kernel}
\end{equation}
where $n_i(z)$ is the normalized redshift distribution of the $i$th bin (see the solid curves in Figure~\ref{fig:z-dist}), and $b(z)$ is the linear galaxy clustering bias, representing the connection between clustering of galaxy and the underlying matter density field. We adopt a linear galaxy bias $b(z) = 1 + 0.84z$ \citep{galaxy-bias}, and in our fiducial model, we assume galaxy bias is a constant in each tomography bin, with value of $b^i = 1 + 0.84z_{\text{cen}}^i$, where $z_{\text{cen}}^i$ is the central redshift of a redshift bin. In Equation~(\ref{eq:C_gg-2}), the second term denotes the shot-noise, and $\Bar{n}_i$ is the mean galaxy surface density in the $i$th bin. We find that $\Bar{n}_i=$ 7.9, 11.5, 4.6 and 3.7 arcmin$^{-2}$ for the four tomographic bins. $N_{\text{sys}}^{\text{g}}$ is the systematic noise, which may arise from spatially varying dust extinction, instrumentation effects and so on \citep[][]{sys_gg_1, Zhan06}. We take $N_{\text{sys}}^{\text{g}} = 10^{-8}$, and assume it is independent to tomographic bins or scales \citep{Gong-CSST-2019}.

In order to avoid the non-linear effect, and improve the prediction accuracy of the theoretical model, we set a minimum scale $k_{\rm max}=0.3\ h\,\rm Mpc^{-1}$ to eliminate the non-linear effect \citep{LSST-DESC, Wenzl-2021}. The corresponding multiple scale cuts in different tomographic bins are $\ell_{\text{max}} = \{359, 920, 1316, 1891\}$, which have been shown in dotted vertical lines in Figure~\ref{fig:scale-cut}. We can find that the cut scales are from $\ell\sim300$ to 2000, and the ratio of non-linear to linear power spectra are less than 1.5 in this scale range. It indicates that the current wavenumber cut is good enough to reduce the non-linear effect, and make the prediction of the theoretical power spectrum mainly stay in the linear regime. 

In the left panel of Figure~\ref{fig:shear-with-components}, the signal, shot-noise, and systematical components of the theoretical galaxy angular power spectrum have been shown in blue solid, black dotted, and black dashed curves, respectively. The total power spectrum is in black solid curves. In Figure~\ref{fig:gg_power_spectrum}, we show the theoretical (blue solid curves) and predicated (red data points) galaxy auto and cross angular power spectra for different tomographic bins. The estimation of covariance matrix and error bars for the data points will be discussed at the end of this section. For each data point, we add a random shift from Gaussian distribution which is derived from the covariance matrix. Note that only the two adjacent bins have the cross power spectrum, since there is no overlapping for galaxy redshift distribution in other cases given the redshift scatter we assume. So we totally have seven auto and cross angular power spectra as shown in Figure~\ref{fig:gg_power_spectrum}. To avoid the effect of assuming the Limber approximation, we only use the data at $\ell>50$ (gray region in the left of each panel) \citep{Fang-beyond-limber}. The data at $\ell\le50$ can be used after taking account of the full expression of the power spectrum in a spherical sky in the future analysis. The data in the non-linear region (gray region in the right of each panel) are also excluded in our data fitting process.

\subsection{Cosmic Shear Power Spectrum}\label{subsec:CS}

\begin{figure*}
	\includegraphics[scale = 0.52]{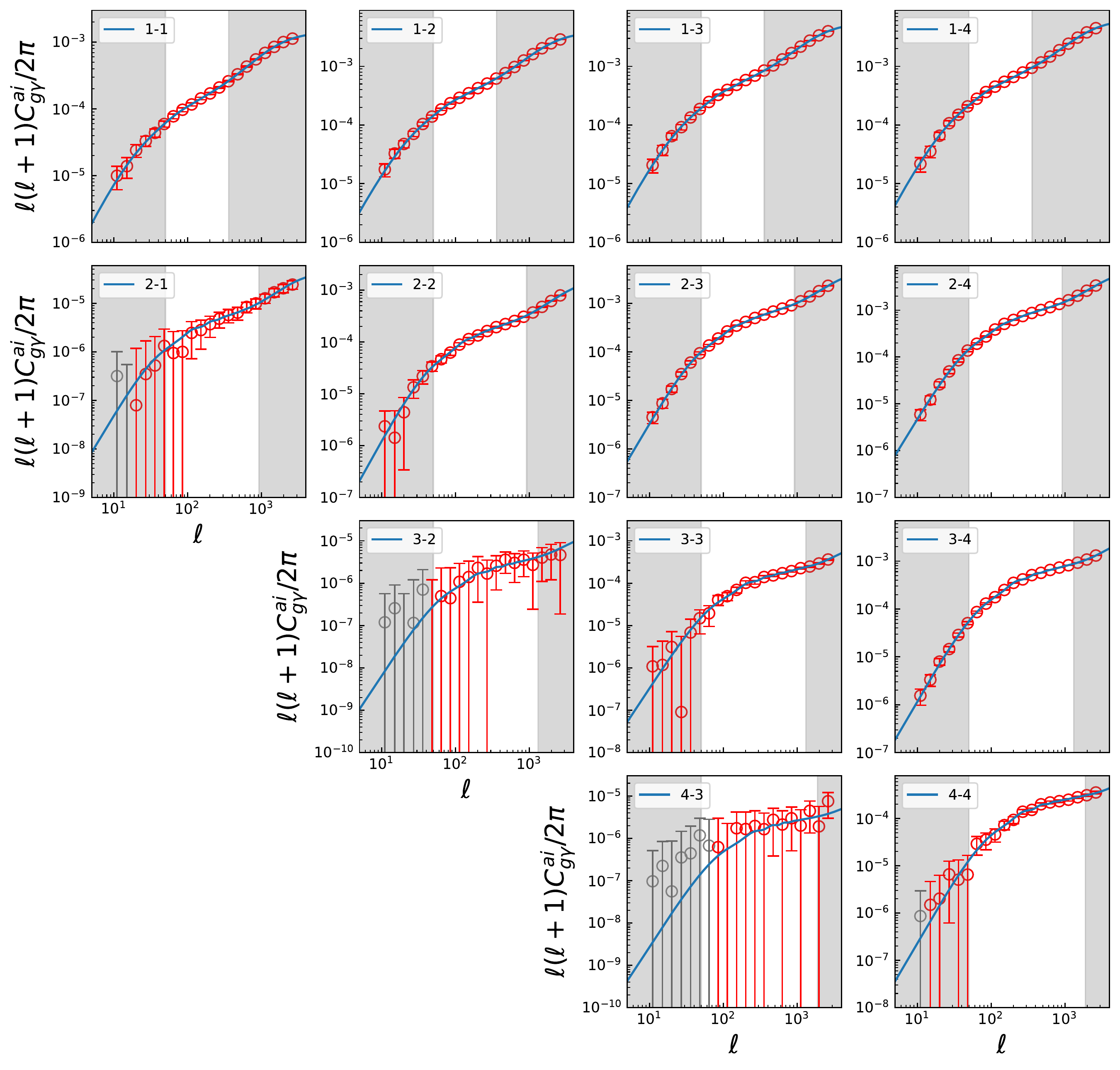}
    \caption{The mock CSST galaxy-galaxy lensing power spectra of the four tomographic bins. The blue curves and red data points denote the theoretical models and mock data. We discard the cross power spectra with low amplitudes. In our cosmological analysis, we only consider the data points with $S/N \ge 1$ (red data points), and the data points with $S/N<1$ (gray data points) have been excluded. The gray regions denote the excluded scales, which are restricted by the galaxy clustering survey.}
    \label{fig:cross_power_spectrum}
\end{figure*}

The observed cosmic shear power spectrum is composed of several components, i.e. signal power spectrum, shape shot-noise, multiplicative, and additive errors. The auto and cross forms in the $i$th and $j$th bins can be expressed as \citep[e.g.][]{Huterer2006} 
\begin{equation}
    \Tilde{C}_{\gamma \gamma}^{ij} = (1+m_i)(1+m_j)C_{\gamma \gamma}^{ij}(\ell) + \delta_{ij}\frac{\sigma_{\gamma}^2}{\Bar{n}_i}+N_{\text{add}}^{\gamma}.
    \label{eq:shear-1}
\end{equation}
Here $m_i$ is the parameter accounting for multiplicative error, which represents one of the main uncertainties in the shear calibration. The second term is the contribution from shape shot noise, where $\sigma_{\gamma} = 0.2$ is due to the intrinsic shape of galaxies and measurement errors. The $N_{\text{add}}^{\gamma}$ is the additive error, including point spread function (PSF), instrumentation effects and so on \citep[][]{sys_shear_1, sys_shear_2}. We assume $N_{\text{add}}^{\gamma} = 10^{-9}$, and it is independent on the tomographic bins or scales \citep{Gong-CSST-2019}. 

The $C_{\gamma \gamma}^{ij}(\ell)$ is the shear signal power spectrum, which has three components \citep{Hildebrandt2017},
\begin{equation}
    C_{\gamma \gamma}^{ij} = P_{\kappa}^{ij}(\ell) + C_{\text{II}}^{ij}(\ell) + C_{\text{GI}}^{ij}(\ell),
    \label{eq:shear-2}
\end{equation}
where $P_{\kappa}^{ij}(\ell)$ is the convergence power spectrum, and $C_{\text{II}}^{ij}(\ell)$ and $C_{\text{GI}}^{ij}(\ell)$ are so called Intrinsic-Intrinsic and Gravitational-Intrinsic power spectra \citep{Joachimi2015}. Assuming the Limber approximation \citep{Limber}, the convergence power spectrum can be obtained by directly integrating over the matter power spectrum weighted by lensing kernel \citep{Hu&Jain2004}, which takes the form as
\begin{equation}
    P_{\kappa}^{ij}(\ell) = \frac{1}{c} \int dz H(z)D_{\text{A}}^{-2}(z)W_{\kappa}^{i}(z)W_{\kappa}^{j}(z)
    P_{\text{m}}\left(\frac{\ell + 1/2}{D_{\text{A}}(z)}, z\right).
    \label{eq:shear-3}
\end{equation}
The lensing kernel in the $i$th tomographic bin is given by
\begin{equation}
    W_{\kappa}^i(z) = \frac{3\Omega_{\rm m} H_0^2}{2H(z)c}\frac{D_{\text{A}}(z)}{a} \int_z^{\infty} dz^{\prime} n_i(z^{\prime}) \frac{D_{\text{A}}(z, z^{\prime})}{D_{\text{A}}(z^{\prime})}.
    \label{eq:shear_kernel}
\end{equation}
In the right panel of Figure~\ref{fig:z-dist}, we show the lensing kernel for the four tomographic bins. We can see that the weighting kernels in high-redshift tomographic bins have relatively wide distributions, which can cover the ranges of the kernels in lower redshift bins.

The intrinsic-intrinsic power spectrum represents the auto-correlation of the intrinsic shape between the neighboring galaxies or cross-correlation between galaxies from different tomographic bins. It is given by
\begin{equation}
    C_{\text{II}}^{ij} = \frac{1}{c} \int dz H(z)D_{\text{A}}^{-2}(z)F^i(z)F^j(z)P_{\text{m}}\left(\frac{\ell + 1/2}{D_{\text{A}}(z)}, z\right).
    \label{eq:C_II}
\end{equation}
Here the weighted kernel of the intrinsic alignment effect can be written as \citep{Hildebrandt2017}
\begin{equation}
    F^i(z) = A_{\text{IA}}C_1 \rho_{\text{c}} \frac{\Omega_{\text{m}}}{D(z)} \frac{n_i(z)}{\Bar{n}_i} \left( \frac{1+z}{1+z_0} \right)^{\eta_{\text{IA}}} \left( \frac{L_i}{L_0} \right)^{\beta_{\text{IA}}},
    \label{eq:IA_kernel}
\end{equation}
where $C_1 = 5 \times 10^{-14} h^{-2}$ $\text{M}_{\odot}^{-1}\text{Mpc}^3$ is a constant, $\rho_{\text{c}}$ is the critical density of present day, $D(z)$ is the linear growth factor which is normalized to unity at $z = 0$. $A_{\text{IA}}$, $\eta_{\text{IA}}$ and $\beta_{\text{IA}}$ are free parameters, and $z_0$ and $L_0$ are pivot redshift and luminosity. For simplicity, we would not consider the luminosity dependence here, and fix $\beta_\text{{IA}} = 0$. Then we set $z_0 = 0.6$, and take of $A_{\text{IA}}=1$ and $\eta_{\text{IA}}=0$ as fiducial values.
On the other hand, the gravitational-intrinsic power spectrum contains two parts. The first one denotes the correlation between the intrinsic shapes of foreground galaxies and the cosmic shears of the background galaxies, and the second one is between foreground shear and background shape (which has much lower signal than the first one). It can be written by
\begin{equation}
\begin{split}
    C_{\text{GI}}^{ij} = \frac{1}{c} \int dz H(z)D_{\text{A}}^{-2}(z) [ W^i(z)F^j(z)\\
    + W^j(z)F^i(z) ] P_{\text{m}}\left(\frac{\ell + 1/2}{D_{\text{A}}(z)}, z\right).
    \label{eq:C_GI}
\end{split}
\end{equation}

In the middle panel of Figure~\ref{fig:shear-with-components}, we show the components of the observed shear power spectrum, including convergence power spectrum, intrinsic alignment terms, shape shot-noise term, and additive error term. In Figure~\ref{fig:shear_power_spectrum}, the ten predicated observed shear power spectra from the CSST photometric survey for four tomographic bins have been shown. Unlike the galaxy clustering case, we can obtain all of the cross power spectra for different two bins, since the lensing kernels have wide redshift distributions. Besides, since the cosmic shear can trace the total underlying matter field and can precisely measure the matter fluctuations at relatively small scales, we use the data in the scale range $50<\ell<3000$ \citep{Fang-beyond-limber}, considering Limber approximation effect.

\subsection{Galaxy-Galaxy Lensing Correlation}\label{subsec:GGL}

The galaxy-galaxy lensing power spectrum is the cross-correlation between galaxy clustering and cosmic shear signal, and it contains two components \citep{Abbott2018DES},
\begin{equation}
    C_{\text{g}\gamma}^{ai}(\ell) = C_{\text{g}\kappa}^{ai}(\ell) + C_{\text{gI}}^{ai}(\ell).
    \label{eq:ggl-1}
\end{equation}
Here we use $a$ and $b$ to denote galaxy samples, and $i$ and $j$ denote cosmic shear samples.  
The first term $C_{\text{g}\kappa}^{ai}(\ell)$ stands for the cross-correlation between galaxy clustering and cosmic shear, which is given by
\begin{equation}
    C_{\text{g}\kappa}^{ai}(\ell) = \frac{1}{c} \int dz H(z)D_{\text{A}}^{-2}(z)W_{\text{g}}^{a}(z)W_{\kappa}^{i}(z)P_{\text{m}}\left(\frac{\ell + 1/2}{D_{\text{A}}(z)}, z\right).
    \label{eq:ggl-2}
\end{equation}
The second term $C_{\text{gI}}^{ai}(\ell)$ represents the cross-correlation between galaxy clustering and intrinsic alignment,
\begin{equation}
   C_{\rm gI}^{ai}(\ell) =  \frac{1}{c} \int dz H(z)D_{\rm A}^{-2}(z)W_{\text{g}}^{a}(z)F^{i}(z)P_{\text{m}}\left( \frac{\ell + 1/2}{D_{\text{A}}(z)}, z \right).
    \label{eq:ggl-3}
\end{equation}

By comparing the left and right panels of Figure~\ref{fig:z-dist}, we can see that the lensing kernel (the right panel) of high redshift has rather large overlap with galaxy clustering kernel (the left panel) of low redshift. On the contrary, low redshift lensing kernel has less overlap with high redshift galaxy clustering kernel. So we expect significant signal for $C_{\text{g}\gamma}^{ai}$ when $a < i$, and a very low amplitude for $C_{\text{g}\gamma}^{ai}$ for $a > i$. It is also consistent with the physics instincts, that we expect a strong correlation between the background shear and the foreground galaxy clustering, since the galaxy clustering trace the underlying matter distributions which play the role of bending the light from the background galaxies. On the other hand, the low redshift shear signal has less correlation with the background matter distribution. Therefore, not all cross galaxy-galaxy lensing power spectra between different tomographic bins have significant and detectable signals. We show the galaxy-galaxy lensing power spectra in Figure~\ref{fig:cross_power_spectrum}. We totally have 13 galaxy-galaxy lensing power spectra for the four tomographic bins. The minimum and maximum multiple scales we consider in the fitting process are restricted by the photometric galaxy clustering survey.

\subsection{Covariance Matrix}\label{subsec:CM}

In order to estimate the constraint power of the probes, we need to calculate the covariance matrix. For the angular galaxy and cosmic shear power spectra, we assume the main contribution is from the Gaussian covariance and the non-Gaussian terms can be ignored \citep{Hu&Jain2004}, and then we have
\begin{equation}
\begin{aligned}
    &{\rm Cov}\left[ \Tilde{C}_{\text{XY}}^{ij}(\ell), \Tilde{C}_{\text{XY}}^{mn}(\ell^{\prime}) \right] \\
    &= \frac{\delta_{\ell \ell ^{\prime}}}{f_{\text{sky}}\Delta\ell(2\ell+1)}
    \left[ 
    \Tilde{C}_{\text{XY}}^{im}(\ell) \Tilde{C}_{\text{XY}}^{jn}(\ell) + 
    \Tilde{C}_{\text{XY}}^{in}(\ell) \Tilde{C}_{\text{XY}}^{jm}(\ell) 
    \right],
    \label{eq:cov-1}
\end{aligned}
\end{equation}
where $f_{\rm sky}$ is the sky fraction, and it is about 42\% sky coverage for the CSST survey. Here the label of bin pair is commutable, $C_{\text{XY}}^{ij} = C_{\text{XY}}^{ji}$ (where $\text{XY} = \{\gamma \gamma, \text{gg}\}$). So the covariance matrix should be a 10$\times$10 matrix in the four-tomographic-bin case. After removing the elements of low-amplitude cross power spectra in the galaxy clustering survey, it is a $7\times7$ matrix for the galaxy angular power spectrum. In the galaxy-galaxy lensing correlation, the label of bin pair is non-commutable, i.e. $C_{\text{g}\gamma}^{ij} \neq C_{\text{g}\gamma}^{ji}$. The Gauassian covariance matrix of galaxy-galaxy lensing can be expressed as
\begin{equation}
\begin{aligned}
     &{\rm Cov}\left[ C_{\text{g}\gamma}^{ai}(\ell), C_{\text{g}\gamma}^{bj}(\ell^{\prime}) \right]\\
     &=  \frac{\delta_{\ell \ell ^{\prime}}}{f_{\text{sky}}\Delta\ell(2\ell+1)}
     \left[
     \Tilde{C}_{\text{gg}}^{ab}(\ell) \Tilde{C}_{\gamma \gamma}^{ij}(\ell)
     + C_{\text{g}\gamma}^{aj}(\ell) C_{\text{g}\gamma}^{bi}(\ell)
     \right].
    \label{eq:cov-2}
\end{aligned}
\end{equation}
Here the covariance matrix should be a $16\times16$ matrix for the four tomographic bins, and it becomes a $13\times13$ matrix after removing the elements of low-amplitude power spectra. The error bar of each data point shown in Figure~\ref{fig:gg_power_spectrum}-\ref{fig:cross_power_spectrum} is derived from the square root of the diagonal elements of the covariance matrix.

\begin{table}
    \centering
    \begin{tabular}{lcccccr}
       \hline
       \text{Parameter} & \text{Fiducial Value} & \text{Prior}  \\
        
       \hline
       \text{Cosmological Parameter} \\
       \hline
       $\Omega_{\text{m}}$ & 0.32 & flat (0, 0.6)  \\
       $\Omega_{\text{b}}$ & 0.048 & flat (0.01, 0.1) \\
       h & 0.6774 & flat (0.4, 1.0)  \\
       $n_{\text{s}}$ & 0.96 & flat (0.8, 1.2)  \\
       $w$ & -1 & flat(-2, 0)  \\
       $\Sigma m_{\nu}$ & 0.06 & flat (0, 2) \\
       $\sigma_8$ & 0.8 & flat (0.4, 1.2)  \\
       \hline
       \text{Intrinsic Alignment} \\
       \hline
       $A_{\text{IA}}$ & 1 & flat (-5, 5)  \\
       $\eta_{\text{IA}}$ & 0 & flat (-5, 5) \\
       \hline
       \text{Baryonic Effect} \\
       \hline
       $\text{log}_{10}(T_{\text{AGN}}/\text{K})$ & 7.8 & flat (7.4, 8.3)   \\
       \hline
       \text{Galaxy Bias} \\
       \hline
       $b^i$ & (1.252,1.756,2.26,3.436) & flat (0, 5)  \\
       \hline
       \text{Photo-z Bias} \\
       \hline
       $\Delta z^i$ & (0,0,0,0) & flat (-0.1, 0.1)  \\
       $\sigma_z^i/\sigma_{z,\text{fid}}^i$ ($\sigma_{z,\text{fid}}=0.05$) & (1,1,1,1) & flat (0.5, 1.5)\\
       \hline
       \text{Shear Calibration} \\
       \hline
       $m_i$ & (0,0,0,0) & flat (-0.1, 0.1) \\
       \hline
    \end{tabular}
    \caption{Free parameters considered in the constraint process. The first column shows the names of our 26 free parameters. The second and third columns show the fiducial values and the prior ranges of the parameters.}
    \label{tab:params-1}
\end{table}

\begin{figure*}
	\includegraphics[scale = 1.4]{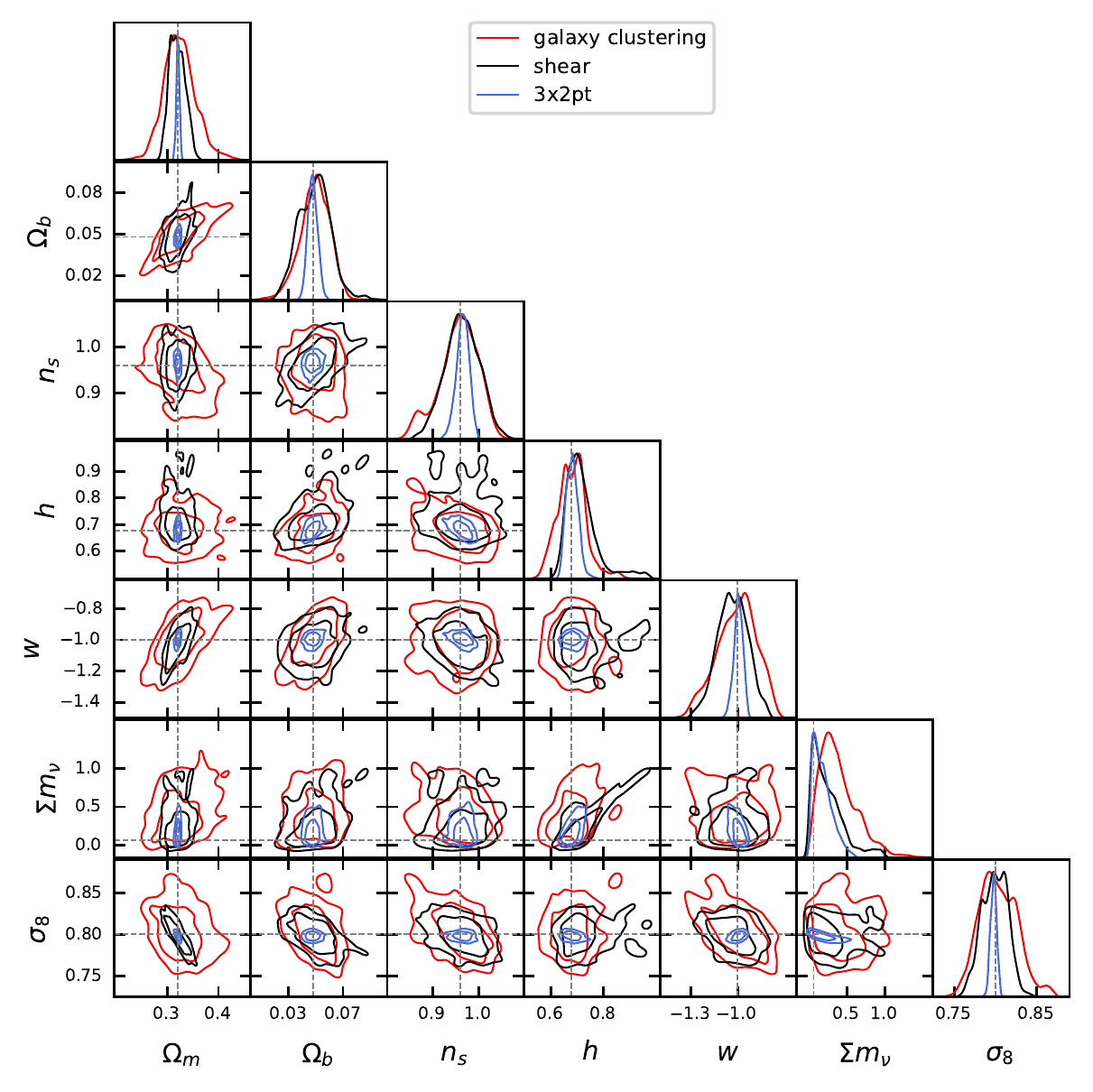}
    \caption{The contour maps of the seven cosmological parameters with 68$\%$ and 95$\%$ confidence levels for the CSST galaxy clustering (red), cosmic shear (black), and 3$\times$2pt (blue) surveys. The 1D PDF of each parameter is also shown. The gray vertical and horizontal dash lines stand for the fiducial values of these parameters.}
    \label{fig:constraint-1}
\end{figure*}

\begin{figure}
	\includegraphics[scale = 0.67]{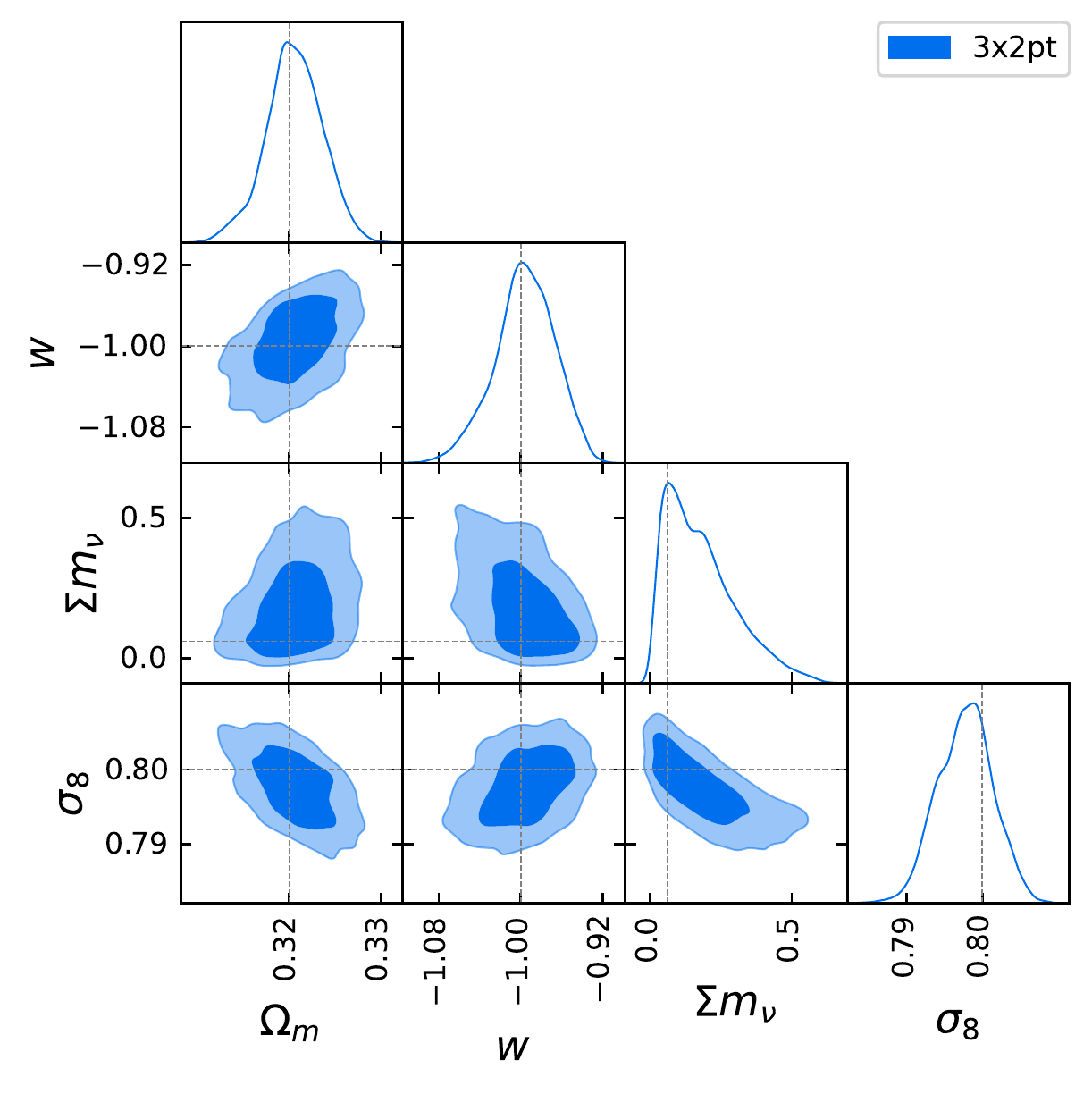}
    \caption{The contour maps of the parameters for dark matter ($\Omega_{\rm m}$ and $\sigma_8$), dark energy ($w$), and total neutrino mass ($\sum m_{\nu}$) for the CSST 3$\times$2pt surveys. The fiducial values are marked by gray vertical and horizontal dash lines.}
    \label{fig:constraint-2}
\end{figure}

\begin{figure}
	\includegraphics[scale = 0.67]{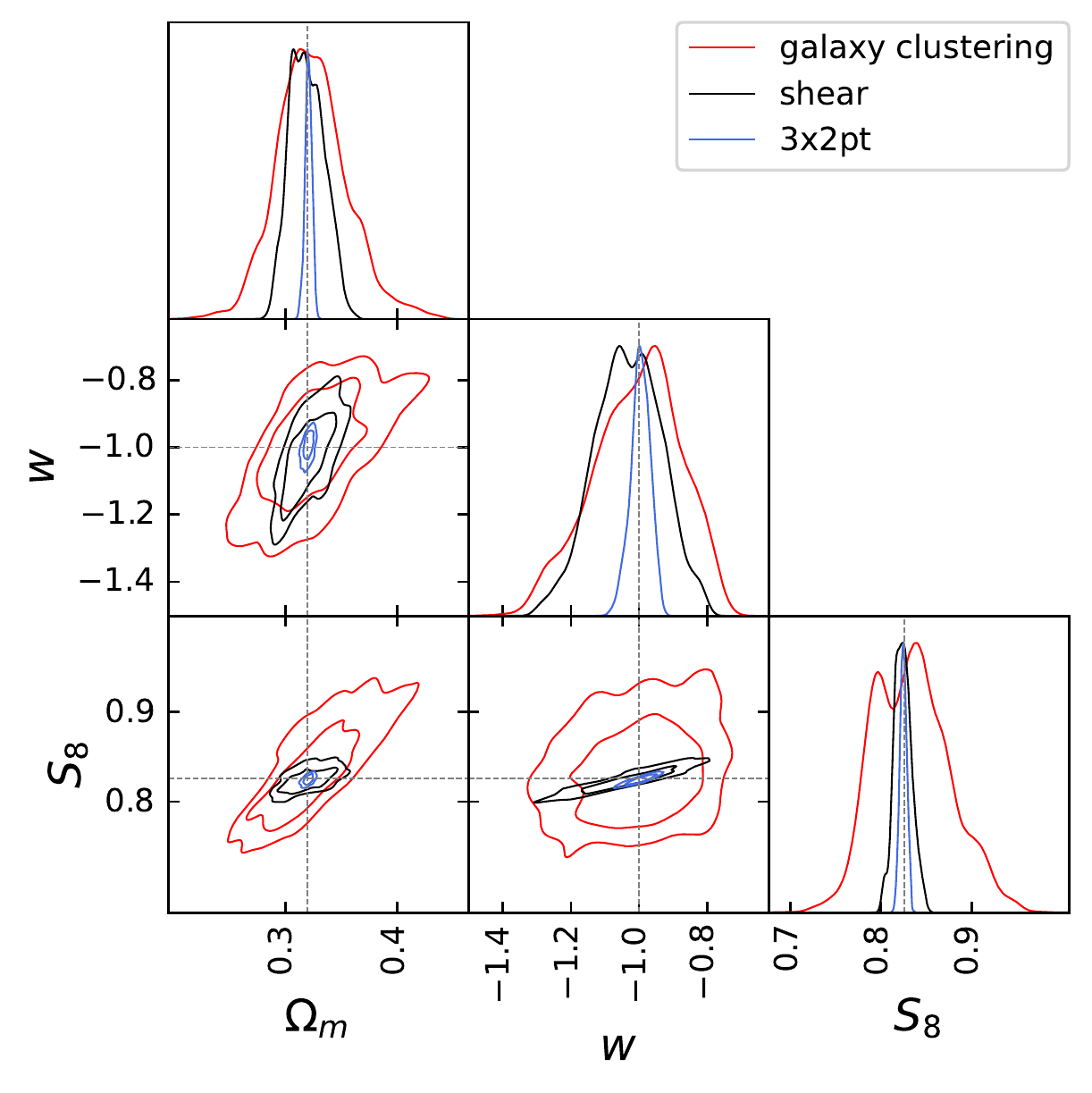}
    \caption{The constraints on $S_8$, $w$ and $\Omega_{\text{m}}$ from the CSST galaxy clustering (red), cosmic shear(black), and 3$\times$2pt (blue) surveys. The gray vertical and horizontal dash lines denote the fiducial values.}
    \label{fig:constraint-3}
\end{figure}

\section{Constraint and results}\label{sec:results}

\subsection{Fitting Method}\label{subsec:FM}

The $\chi^2$ method is adopted to fit the mock data of the CSST photometric surveys, which takes the form as
\begin{equation}
    \chi^2 = \sum_{\ell_{\text{min}}}^{\ell_{\text{max}}}\left[ 
    \vec{d}(\ell) - \vec{t}(\ell) \right] 
    {\rm Cov}^{-1} \left[ \vec{d}(\ell), \vec{d}(\ell^{\prime}) \right]
    \left[ \vec{d}(\ell) - \vec{t}(\ell) \right],
    \label{eq:chi2}
\end{equation}
where $\vec{d}(\ell)$ and $\vec{t}(\ell)$ are the observed and theoretical data vectors, respectively. Then we can calculate the likelihood function as $\mathcal{L} \propto {\rm exp}(-\chi^2/2)$. The total chi-square for the 3$\times$2pt data can be estimated by $\chi^2_{\rm tot} = \chi^2_{\rm gg} + \chi^2_{\rm \gamma\gamma} + \chi^2_{\rm g\gamma}$, where $\chi^2_{\rm gg}$, $\chi^2_{\rm \gamma\gamma}$, and $\chi^2_{\rm g\gamma}$ are the chi-squares for the photometric galaxy clustering, cosmic shear, and galaxy-galaxy lensing data. 

We make use of the {\tt emcee} package \citep{emcee}, which is a Markov Chain Monte Carlo (MCMC) Ensemble sampler based on the affine-invariant ensemble sampling algorithm \citep{Goodman2010}, to perform the fitting process. We initialize 500 walkers around our fiducial parameters, and obtain 1 million steps. Then we discard the first $30\%$ steps as the burn-in. We summarize the free parameters we include for each survey, and their fiducial values and priors in Table~\ref{tab:params-1}. We include total neutrino mass $\sum m_{\nu}$, dark energy equation of state $w$, reduced Hubble constant $h$, and other four cosmological parameters for constraining. The systematical parameters, such as the ones from galaxy bias, photo-$z$ uncertainty, intrinsic alignment, shear calibration and baryonic feedback, are also considered. Totally we have 7 cosmological parameters, and 19 systematical parameters in the fitting process.

\subsection{Cosmological Parameters}\label{subsec:CP}

\begin{table*}
    \centering
    \begin{tabular}{lcccccr}
       \hline
       \text{Parameter} & \text{Fiducial Value} & \text{Constraints by} & \text{Constraints by} & \text{Constraints by} \\
         & & \text{Galaxy Clustering} & \text{Weak Lensing} & \text{3$\times$2pt}\\
       \hline
       \text{Cosmological Parameter} \\
       \hline
       $\Omega_{\text{m}}$ & 0.32 & $0.324^{+0.026}_{-0.034}$ ($9.26\%$) & $0.319^{+0.014}_{-0.018}$ ($5.02\%$) & $0.3206^{+0.0031}_{-0.0031}$ ($0.97\%$) \\
       $\Omega_{\text{b}}$ & 0.048 & $0.0494^{+0.012}_{-0.0088} ($21.05\%$)$ & $0.049^{+0.012}_{-0.012}$ ($24.49\%$) & $0.0478^{+0.0032}_{-0.0039}$ ($7.43\%$) \\
       h & 0.6774 & $0.680^{+0.056}_{-0.056}$ ($8.24\%$) & $0.718^{+0.025}_{-0.067}$ ($6.41\%$) & $0.681^{+0.020}_{-0.027}$ ($3.45\%$) \\
       $n_{\text{s}}$ & 0.96 & $0.955^{+0.051}_{-0.030}$ ($4.24\%$) & $0.964^{+0.040}_{-0.035}$ ($3.89\%$) & $0.965^{+0.015}_{-0.012}$ ($1.40\%$) \\
       $w$ & -1 & $-1.01^{+0.14}_{-0.12}$ ($12.87\%$) & $-1.032^{+0.098}_{-0.098}$ ($9.50\%$) & $-0.996^{+0.032}_{-0.025}$ ($2.86\%$) \\
       $\Sigma m_{\nu}$ & 0.06 & $< 0.708 \text{ eV}$ & $< 0.481 \text{ eV}$ & $< 0.36 \text{ eV}$\\
       $\sigma_8$ & 0.8 & $0.804^{+0.023}_{-0.023} ($2.86\%$) $ & $0.799^{+0.016}_{-0.016}$ ($2.00\%$) & $0.7977^{+0.0037}_{-0.0037}$ ($0.46\%$) \\
       \hline
       \text{Intrinsic Alignment} \\
       \hline
       $A_{\text{IA}}$ & 1 & --- & $1.004^{+0.062}_{-0.073}$ ($6.72\%$) & $0.993^{+0.018}_{-0.018}$ ($1.81\%$) \\
       $\eta_{\text{IA}}$ & 0& --- & $-0.07^{+0.29}_{-0.34}$ ($231\%$) & $0.056^{+0.038}_{-0.033}$ ($63.39\%$) \\
       \hline
       \text{Baryonic Effect} \\
       \hline
       $\text{log}_{10}(T_{\text{AGN}}/\text{K})$ & 7.8 & $7.80^{+0.10}_{-0.10}$ ($1.28\%$) & $7.805^{+0.066}_{-0.13}$ ($1.26\%$) & $7.748^{+0.033}_{-0.033}$ ($0.43\%$)\\
       \hline
    \end{tabular}
    \caption{The constraint result of the 7 cosmological parameters, 2 intrinsic alignment parameters and one baryonic parameter in our model for the CSST galaxy clustering, weak lensing, and 3$\times$2pt surveys. The constraint relative accuracy for each parameter is also shown in the bracket.}
    \label{tab:params-2}
\end{table*}

In Figure~\ref{fig:constraint-1}, we show the marginalized contours for our 7 cosmological parameters from galaxy clustering only, weak lensing only, and the joint 3$\times$2pt constraint. We show the 68$\%$ and 95$\%$ confidence levels (C.L.) contours, and the marginalized 1D posteriors. The gray dash line mark the fiducial values of the parameters. And we show the main constraint results of our 7 cosmological parameters, 2 intrinsic alignment parameters and one baryonic parameter from the three CSST probes in 68$\%$ confidence level in Table~\ref{tab:params-2}. The percent constraint results or constraint relative accuracies for the parameters are also shown in the brackets of the table.

For the CSST photometric galaxy clustering survey, we obtain a 9.26$\%$ constraint on $\Omega_{\text{m}}$, 12.87$\%$ constraint on $w$ and 2.86$\%$ constrant on $\sigma_8$ in 1$\sigma$ confidence level. It is a great improvement compared to the current 2D galaxy clustering surveys. For example, our result is better than the DES Y3 $\omega (\theta) + \gamma_{t}$ result \citep{DES-Y3-gg} by a factor of 1.5 $\sim$ 3, although they are using a joint fit of angular galaxy clustering and galaxy-galaxy lensing. Due to deeper magnitude limit, wider survey area, and larger wavelength coverage with seven photometric bands, CSST can obtain a much larger photometric galaxy sample with lower Poisson noise and  better accuracy of photo-z calibration, which can make CSST obtain better constraint results with angular galaxy clustering data only. We also compare our result with spectroscopic survey, for example the extended Baryon Oscillation Spectroscopic Survey (eBOSS) cosmological analysis \citep{alam2021eBOSS}. They are using spectroscopic data to study the 3D galaxy clustering, which contains more information about the structure growth than the 2D clustering that we use, since they can achieve much higher accuracy on redshift measurement. However, limited by much lower galaxy density that leads to relatively large Poisson noise in the eBOSS survey, we still can achieve similar constraint power compared to their result. Although it may be not quite suitable to directly compare our photometric result with them, it shows that CSST photometric galaxy clustering survey has a great potential in the cosmological studies. 

For the CSST weak lensing survey, we obtain a 5.02$\%$ constraint on $\Omega_{\text{m}}$,  9.50$\%$ constraint on $w$ and  2.00$\%$ constrant on $\sigma_8$ in 1$\sigma$ confidence level. This result is better than the ongoing DES \citep{DES-Y3-shear} and KiDS \citep{Heymans-KiDS} by a factor $\sim 4$. Since CSST has a high spatial resolution $\sim$0.15'' (80\% energy concentration region) with a relatively regular PSF shape close to the Gaussian profile \citep{Gong-CSST-2019}, it can obtain excellent shape measurements with good imaging quality, which is an advantage in the weak lensing survey. Besides, CSST also has a wide survey area, deep magnitude limit, and large wavelength coverage with seven bands, that can significantly reduce the uncertainties from cosmic variance, shot noise, and photo-$z$ calibration.  All of these advantages make the CSST shear or weak lensing survey be able to achieve an impressive improvement on parameter constraints. This result is also consistent with the previous study of the CSST weak lensing survey given in \citet{Gong-CSST-2019}, although we have considered neutrinos and baryonic effect in this work. We also notice that the constraints from the CSST shear measurements are basically comparable or even better than that from the photometric galaxy clustering survey, since cosmic shear surveys include more information at small-scale non-linear regimes.

When combining the data of the CSST photometric galaxy clustering, weak lensing and galaxy-galaxy lensing surveys, i.e. 3$\times$2pt, we can see that the constraint results have magnificent improvements for all cosmological parameters. Since the contours of 3$\times$2pt are too small in Figure.~\ref{fig:constraint-1}, in particular, we show the constraint results of the four cosmological parameters in Figure.~\ref{fig:constraint-2}, i.e. $\Omega_{\text{m}}$, $\sigma_8$ and $w$ and $\Sigma m_{\nu}$. And the constraint result is 0.97$\%$ constraint on $\Omega_{\text{m}}$, 2.86$\%$ constraint on $w$ and 0.46$\%$ constrant on $\sigma_8$ (we will discuss the constraint on neutrino mass later, in~\ref{subsec:Neutrino-Mass}). We find that they are improved by at least one order of magnitude than the current 3$\times$2pt results, e.g. the ones given in DES Year 3 result \citep{DES2021}. 


In order to break the degeneracy between $\sigma_8$ and $\Omega_{\text{m}}$, we explore the constraint of a less degenerate parameter $S_8 = \sigma_8 (\Omega_{\text{m}}/0.3)^{0.5}$. In Figure~\ref{fig:constraint-3}, we can see that, the contours of weak lensing are horizontal in the $\Omega_{\text{m}}-S_8$ plane now. But for galaxy clustering, the degeneracy between $S_8$ and $\Omega_{\text{m}}$ still remain. In order to break the remain degeneracy, one can set a new parameters $\Sigma_{8} = \sigma_{8}(\Omega_{\text{8}}/0.3)^{\alpha}$, and fit $\alpha$ to capture a best $\Sigma_{8}$ which is perpendicular to the degeneracy between $\sigma_{8}$ and $\Omega_{\text{m}}$ \citep[see e.g.][]{KiDS-Asgari}. Since it is out of the scope of this work, we leave this for future studies on the real data collected by CSST.

\begin{figure*}
    \includegraphics[scale = 0.47]{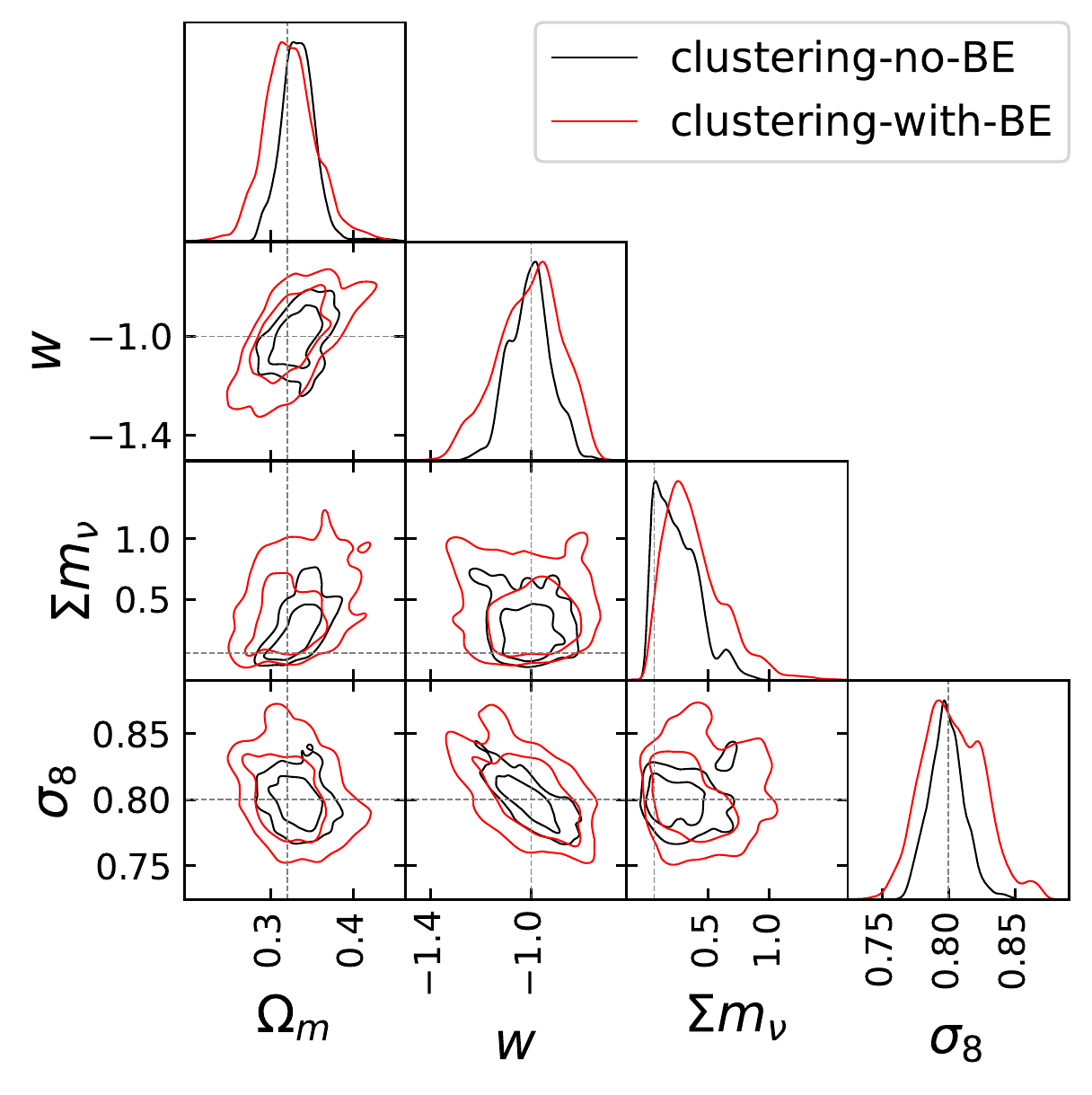}
    \includegraphics[scale = 0.47]{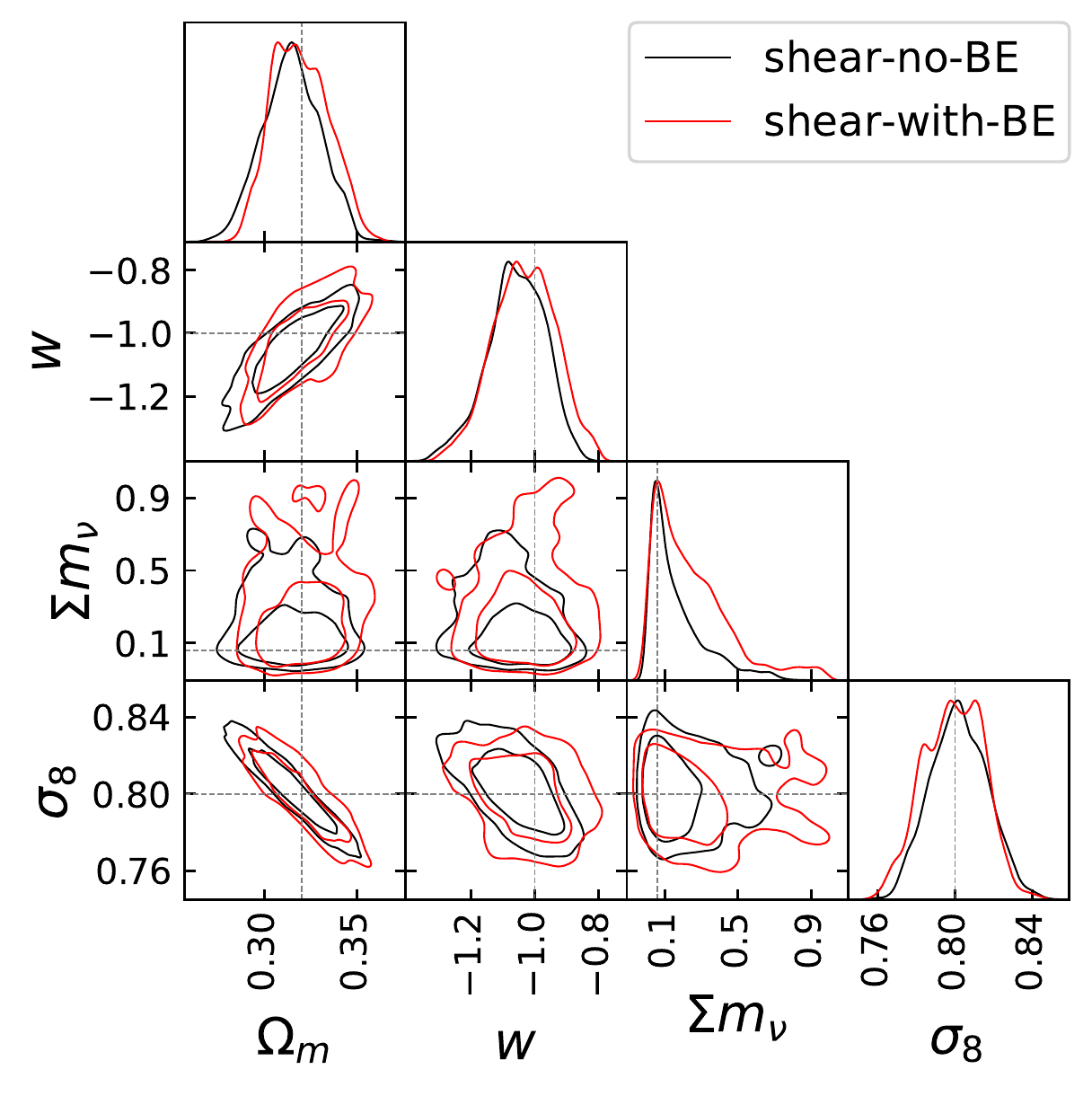}
    \includegraphics[scale = 0.47]{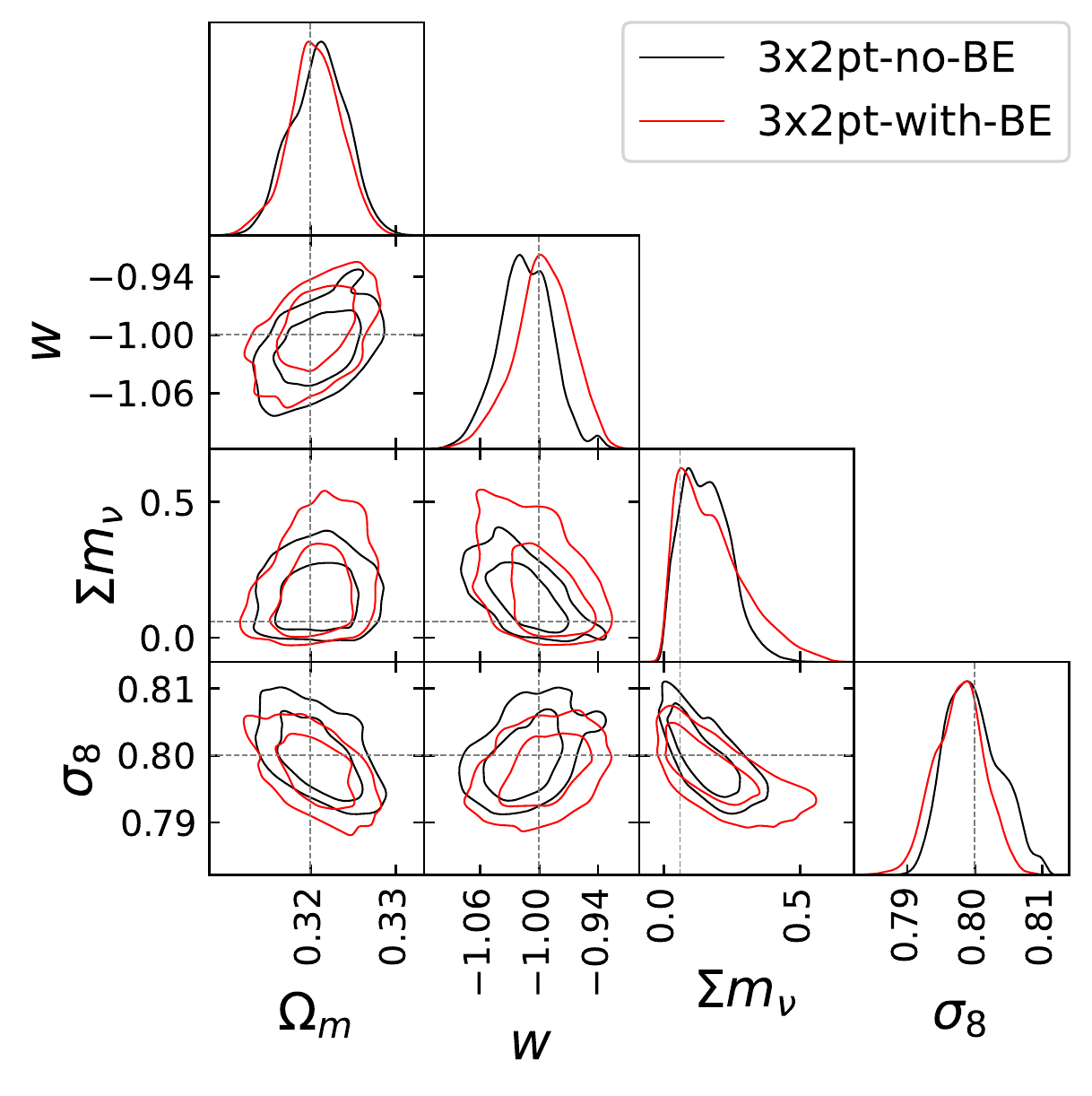}
    \caption{The comparison of the constraint results on $\Omega_{\text{m}}$, $w$, $\sigma_{8}$ and $\Sigma m_{\nu}$ with (red) and without (black) baryonic effect (BE) considered in the CSST photometric surveys. The left, middle and right panel show the results of galaxy clustering, weak lensing, and the joint 3$\times$2pt surveys, respectively.}
    \label{fig:NB-vs-WB}
\end{figure*}

\begin{figure}
    \includegraphics[scale = 0.6]{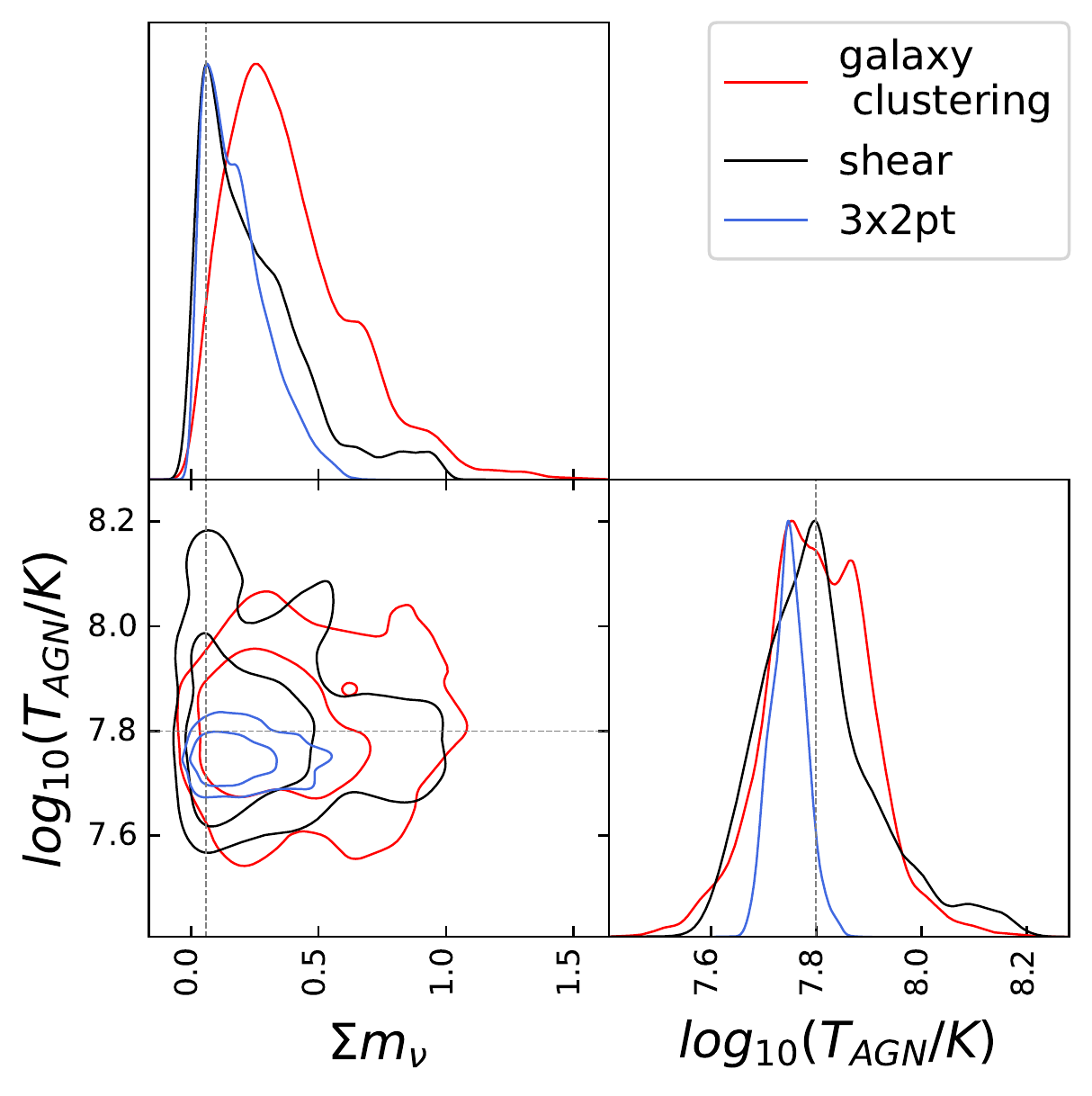}
    \caption{The marginalized constraint results of the baryonic feedback strength parameter $\text{log}_{10}(T_{\text{AGN}}/\text{K})$ and neutrino masses $\Sigma m_{\nu}$. The red, black and blue curves show the results of the CSST galaxy clustering only, weak lensing only, and joint 3$\times$2pt surveys, respectively. The grey dash lines indicate the fiducial values of the paramters in our model.}
    \label{fig:T_AGN}
\end{figure}

\subsection{Neutrino Mass and Impact of Baryons}\label{subsec:Neutrino-Mass}

Because weak lensing is sensitive to small scales physics, it provides a powerful probe to constrain neutrino mass. In our work, we assume normal hierarchy with the sum of neutrino mass equal to the lower limit given by the current oscillation experiments \citep{Amsler2008}. Then the mass of three eigenstates are $m_1 \approx m_2 \approx 0$ eV, and $m_3 \approx 0.06$ eV. Since baryons can impact the matter power spectrum on small scales, it will lead to a significant degeneracy between baryonic parameter and neutrino mass. So we analyze them together in this section.

In Figure~\ref{fig:constraint-1} or Figure~\ref{fig:constraint-2}, the joint 3$\times$2pt surveys can provide upper limits of $\Sigma m_{\nu} < 0.36$ and 0.56 eV for 68\% and 95\% C.L., respectively. Although it is looser than the recent {\it Planck} result \citep{Planck2018-VI} with  $\Sigma m_{\nu} < 0.24$ eV (95$\%$, $\it Planck$ TT, TE, EE+lowE+lensing), which is the most accurate result by far for a survey, our result is actually an extraordinarily strong constraint for an optical galaxy survey. For example, the recent DES Year 3 analysis \citep{DES2021} can not effectively constrain neutrino mass, whether alone or in combination with external BAO, RSD, and supernova data. After combining with $\it Planck$ data, although they can obtain $\Sigma m_{\nu} < 0.13$ eV (95$\%$ C.L.), clearly the constraint power is mainly contributed by the $\it Planck$ data. 

For comparison, we also investigate the constraint results without considering the baryonic effect for galaxy clustering only, shear only and joint 3$\times$2pt surveys, and only the results for $\Omega_{\rm m}$, $\sigma_{8}$, $w$, and $\Sigma m_{\nu}$ have been shown in Figure.~\ref{fig:NB-vs-WB}. As can be seen, all of parameter contours become smaller than the case considering the baryonic effect, since statistically speaking less parameters are included without the baryonic effect. Especially, the errors of neutrino mass almost shrink about 1$\sigma$ for all of the three datasets. This gives upper limits of $\Sigma m_{\nu} < 0.23$ and 0.29 eV at 68\% and 95\% C.L., respectively, for the 3$\times$2pt survey, respectively, which is comparable to the {\it Planck} result. 
Hence, the CSST photometric survey would be a promising survey for exploring the neutrino mass in the future. And if we consider the synergies between the CSST survey and the CMB-S4 (such as the Simons Observatory, \citep{SO-CMB-S4}), we may have a great chance to finally determinate the neutrino mass and solve the hierarchy problem as well.

We also show the constraint results on the AGN temperature parameter $\text{log}_{10}(T_{\text{AGN}}/\text{K})$, which is related to the baryonic effect, in Figure~\ref{fig:T_AGN}. We can see that the constraint from galaxy clustering and weak lensing only are relatively weak, but if we combine them with galaxy-galaxy lensing, the 3$\times$2pt probe can gain a much tighter result. The weak constraint power from the galaxy clustering is probably due to that we cut off many modes on small scales to avoid the uncertainty from the non-linear effect. And in the weak lensing case, since we save the small scales modes up to $\ell = 3000$, the weak lensing survey can probe more small-scale physics. Although weak lensing can not distinguish dark matter and baryon, the baryon can alter the dark matter distribution via gravitational interaction. This eventually changes the total matter distribution, leaving an indirect effect on the weak lensing signal and making the weak lensing survey obtain a better constraint on baryon physics than the galaxy clustering survey. We also show the marginalized constraint results of $\text{log}_{10}(T_{\text{AGN}}/\text{K})$ and $\Sigma m_{\nu}$ in Figure~\ref{fig:T_AGN} for the CSST galaxy clustering only, weak lensing only, and joint 3$\times$2pt surveys, respectively. We find that, there is a relatively weak degeneracy between $\text{log}_{10}(T_{\text{AGN}}/\text{K})$ and the neutrino mass. This result is consistent with the result of \citet{Mead-2021}, where they find the suppression on the power spectrum caused by the baryonic feedback is weakly dependent on the neutrino mass in the {\it WMAP} 9 simulations.

The other systematical parameters of galaxy clustering and weak lensing observations are also jointly constrained in our model, such as galaxy bias, photo-$z$ uncertainties, intrinsic alignment and shear calibration. We show and discuss the constraint results of these parameters in the Appendix. We can find that, although the systematical parameters can degenerate with the cosmological parameters and make the shape of parameter probability space irregular, basically they can still be stringently and correctly constrained in the CSST photometric surveys.

\section{Conclusions}\label{sec:Conclusions}

In this work, we forecast the constraints on the neutrino mass and other cosmological parameters under the $w$CDM model for the CSST photometric surveys. The galaxy clustering, cosmic shear, and galaxy-galaxy lensing data are considered, and the systematics from galaxy bias, intrinsic alignment, photo-$z$ uncertainties, shear calibration, baryonic feedback, and instruments have been included in the analysis.

We generate the mock data based on the COSMOS catalog to obtain galaxy redshift distribution, galaxy surface density, and other necessary information. The MCMC technique is employed in the fitting process. We obtain very tight constraints on the key cosmological parameters, such as $\Omega_{\text{m}}$, $\sigma_8$, and $w$. Our result achieve a factor of $4 \sim 10$ improvement to the ongoing similar optical survey, such as DES and KiDS. We also obtain a stringent constraints on the sum of neutrino mass, that gives $\Sigma m_{\nu} < 0.36$ and 0.56 eV (68\% and 95\% C.L., with baryonic effect for 3$\times$2pt) or $\Sigma m_{\nu} < 0.23$ and 0.29 eV (68\% and 95\% C.L., without baryonic effect for 3$\times$2pt). The constraints on 19 systematical parameters are also explored. We find that the CSST 3$\times$2pt survey is very useful in constraining intrinsic alignment effect and galaxy bias, and also particularly helpful for calibrating the photo-z measurement, which could effectively improve the constraint on the cosmological parameters. We also should note that, although we have considered as many uncertainties and systematics as we can, the results obtained from the real CSST surveys could be worse, since some assumptions made in this work still can be too simple. More realistic results could be derived from the data provided by future observational simulations.

In addition to the 3$\times$2pt surveys, other cosmological probes also can be applied to further improve the constraint results in the CSST surveys. For example, the weak lensing peak counts should significantly tighten the constraints on neutrino mass, since peaks are expected to contain more information of non-linear structures. Besides, the CSST can also perform the 3D spectroscopic galaxy clustering survey, strong gravitational lensing survey, galaxy cluster survey, Type Ia supernova observation, etc. Joint analysis of all these surveys can further help us to explore the unsolved mystery in our Universe, and put extremely stringent constraints on the neutrino mass, properties of dark energy and dark matter, and other important topics. Benefited by its great imaging quality, large survey area, deep magnitude limit, wide wavelength coverage with multiple bands, we expect the CSST photometric survey will be a magnificent space-based sky survey for the cosmological studies.

\section*{Acknowledgements}
H.J.L. and Y.G. acknowledge the support of MOST-2018YFE0120800, 2020SKA0110402, NSFC-11822305, NSFC-11773031, NSFC-11633004, and CAS Interdisciplinary Innovation Team. X.L.C. acknowledges the support of the National Natural Science Foundation of China through grant No. 11473044, 11973047, and the Chinese Academy of Science grants QYZDJ-SSW-SLH017, XDB 23040100, XDA15020200. K.C.C. acknowledges the support from the National Science Foundation of China under the grant 11873102. Z.H.F acknowledges the supports from NSFC of China under No. 11933002 and No. U1931210. This work is also supported by the science research grants from the China Manned Space Project with NO.CMS-CSST-2021-B01 and CMS- CSST-2021-A01.

\section*{Data Availability}

 The data that support the findings of this study are available from the corresponding author, upon reasonable request.



\bibliographystyle{mnras}
\bibliography{ref} 




\appendix

\section{Constraints on systematical parameters}



\begin{figure*}
    \includegraphics[width=\textwidth]{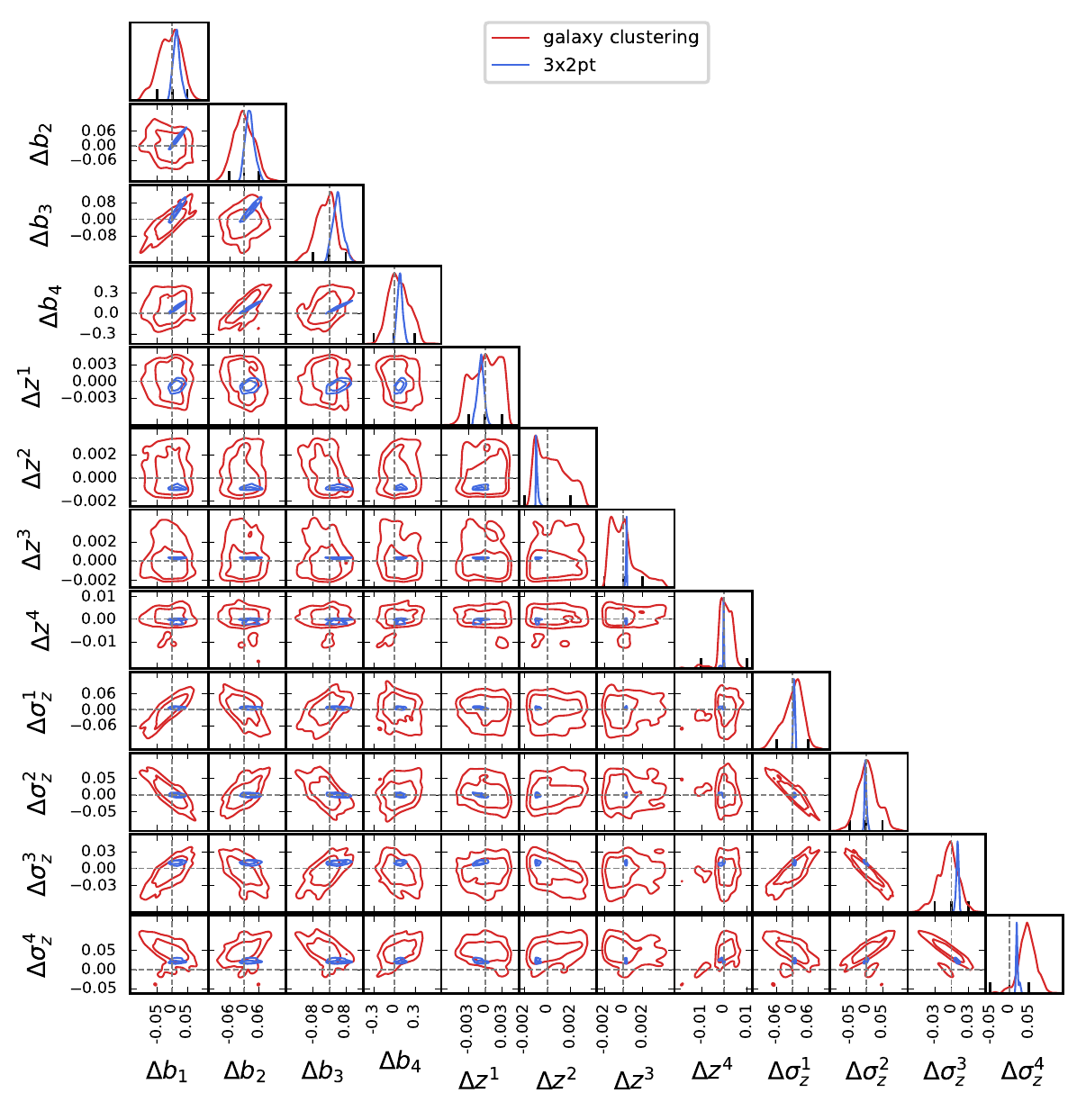}
    \caption{Constraint results of the residuals (i.e. best-fit values minus fiducial values) of the 12 systematical parameters in the CSST galaxy clustering (red) and 3$\times$2pt (blue) surveys. The 68$\%$ and 95$\%$ confidence levels, and 1D PDF for each parameter have been shown.}
    \label{fig:constraint-sys-2}
\end{figure*}

\begin{figure*}
    \includegraphics[width=\textwidth]{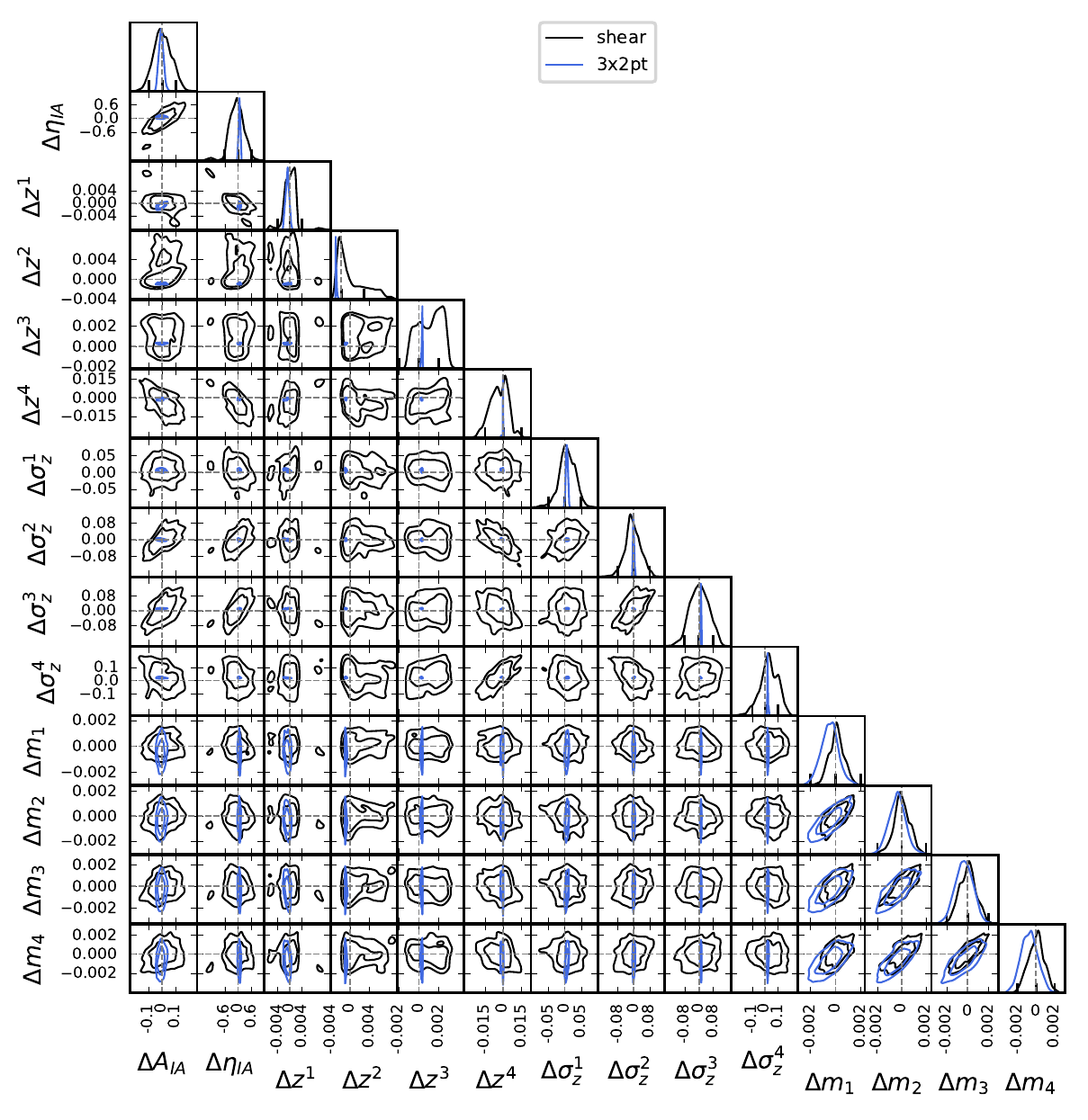}
    \caption{Constraint results of the residuals (i.e. best-fit values minus fiducial values) of the 14 systematical parameters in the CSST cosmic shear (gray) and 3$\times$2pt (blue) surveys. The 68$\%$ and 95$\%$ confidence levels, and 1D PDF for each parameter have been shown.}
    \label{fig:constraint-sys-1}
\end{figure*}

Besides the cosmological parameters, CSST survey also provide constraint on some important systematical parameters relate with survey properties as well. We summarize the constraints on all of these parameters for the CSST galaxy clustering and weak lensing surveys in Figure~\ref{fig:constraint-sys-2} and Figure~\ref{fig:constraint-sys-1}, respectively. For clarity, the constraint on  the residual of each parameter (i.e. best-fit value minus fiducial value) is shown in the two figures.

Galaxy bias is an important parameter in modeling the galaxy clustering. In our 3$\times$2pt analysis, as shown in Figure~\ref{fig:constraint-sys-2}, we obtain $b_1 = 1.2670^{+0.0099}_{-0.014}$, $b_2 = 1.780^{+0.015}_{-0.022}$, $b_3 = 2.301^{+0.023}_{-0.029}$ and $3.519^{+0.035}_{-0.043}$, which are better than the galaxy clustering only result by a factor $\sim 4$. We find that the 1$\sigma$ uncertainty of galaxy bias is within $\pm0.04$, indicating that the CSST survey can put a really tight constraint on the galaxy bias. Note that the linear galaxy bias assumption might break down on small scales (which we have excluded in our analysis). In the future analysis for the real data from galaxy clustering survey, we would consider more complicated model, such as non-linear galaxy bias \citep{Senatore-nonlinear-bias} or galaxy assembly bias \citep{Wechsler-assembly-bias}, and baryonic effect \citep{Lewandowski-baryonic-effect}, and extract more information from small scales. And if we have a better understanding on the galaxy bias, the galaxy clustering survey will become a much more powerful cosmological probe.

In our cosmic shear theoretical model, the intrinsic alignment is assumed to be proportional to $A_{\text{IA}}(1+z)^{\eta_{\text{IA}}}$. In Figure~\ref{fig:constraint-sys-1}, we find that we can obtain stringent constraints on the amplitude with $A_{\text{IA}} = 1.004_{-0.073}^{+0.062}  \text{ (weak lensing only)}$ and $A_{\text{IA}} = 0.993_{-0.018}^{+0.018}  \text{ (3$\times$2pt)}$. But our probes relatively are not sensitive to the power law index, which give $\eta_{\text{IA}} = -0.07^{+0.29}_{-0.34}$ from weak lensing survey and $\eta_{\text{IA}} = 0.056_{-0.033}^{+0.038}$ from 3$\times$2pt. In this work, we didn't separate the intrinsic alignment signal from the total signal, but in some recent work \citep[e.g.][]{Mandelbaum-IA, Tonegawa-IA}, they already measured intrinsic alignment signal from real data. And there are some theoretical works suggest that the intrinsic alignment signal could be a new tool to constrain the cosmological parameters or study dynamical aspects of galaxy evolution models \citep[e.g.][]{Chisari-IA, Taruya-IA}. The CSST survey will have great potential in using the intrinsic alignment as an independent probe, and we will leave this as a future work. For  the parameters of the multiplicative errors, i.e. $m_i$, we find that the constraint power are similar for the weak lensing only and 3$\times$2pt surveys. This indicates that the joint 3$\times$2pt survey may be not quite helpful for significantly improving the shear calibration.

For the parameters of photo-$z$ uncertainty, i.e. the photo-z bias $\Delta z^i$ and photo-z rms parameter $\Delta \sigma_z^i$, we find that the 3$\times$2pt surveys will offer a dramatical improvement compared to galaxy clustering only and weak lensing only surveys. This is because both galaxy clustering and weak lensing surveys are tightly related to the photo-$z$ uncertainties, and the joint survey can provide more information to suppress the photo-z uncertainties and result in excellent constraints on $\Delta z^i$ and $\Delta \sigma_z^i$.


\bsp	
\label{lastpage}
\end{document}